\documentstyle[lscape,epsfig,supertabular] {l-aa}   
\def\phibar{\bar{\varphi}}

\begin{document}
\thesaurus{           
             (20; 13.09.2; 04.03.1; 04.19.1)}  
\title{The first COMPTEL source catalogue}
\author{V.~Sch\"{o}nfelder\inst{1} \and K.~Bennett\inst{4} \and J.J.~Blom\inst{2} \and H.~Bloemen\inst{2} \and W.~Collmar\inst{1} \and A.~Connors\inst{3} \and R.~Diehl\inst{1} \and  W.~Hermsen\inst{2}  
\and A.~Iyudin\inst{1} \and R.M.~Kippen\inst{3} \and  J.~Kn\"odlseder\inst{5} \and L.~Kuiper\inst{2} \and G.G.~Lichti\inst{1} \and M.~McConnell\inst{3} \and  D.~Morris\inst{3}  \and R.~Much\inst{4} \and U.~Oberlack\inst{1} \and J.~Ryan\inst{3} \and G.~Stacy\inst{3} \and H.~Steinle\inst{1} \and A.~Strong\inst{1} \and R. Suleiman\inst{3} \and R.~van~Dijk\inst{4} \and M.~Varendorff\inst{1} \and C.~Winkler\inst{4} \and O.R.~Williams\inst{4}} 
\offprints{\\ \mbox{V.~Sch\"onfelder, vos@mpe.mpg.de}}
\institute{Max-Planck-Institut f\"ur extraterrestrische Physik, 
           D--85740 Garching, Germany
\and 
SRON--Utrecht, Sorbonnelaan 2, NL--3584 CA Utrecht, The Netherlands
\and 
Space Science Center, University of New Hampshire, Durham NH 03824-3525, USA
\and
Astrophysics Division, ESTEC, NL--2200 AG Noordwijk, The Netherlands 
\and
Centre d'Etude Spatiale des Rayonnements (CESR), BP 4346, F-31029 Toulouse Cedex, France
}

\date{Received 28 July 1999; Accepted 20 December 1999}

\maketitle
   
\begin{abstract}
The imaging Compton telescope COMPTEL aboard NASA's Compton Gamma-Ray Observatory has opened the MeV gamma-ray band as a new window to astronomy. COMPTEL provided the first complete all-sky survey in the energy range 0.75 to 30 MeV. The catalogue, presented here, is largely restricted to published results. It contains firm as well as marginal detections of continuum and line emitting sources and presents upper limits for various types of objects. The numbers of the most significant detections are 32 for steady sources and 31 for gamma-ray bursters. Among the continuum sources, detected so far, are spin-down pulsars, stellar black-hole candidates, supernova remnants, interstellar clouds, nuclei of active galaxies, gamma-ray bursters, and the Sun during solar flares. Line detections have been made in the light of the 1.809 MeV $^{26}$Al line, the 1.157 MeV $^{44}$Ti line, the 847 and 1238 keV $^{56}$Co lines, and the neutron capture line at 2.223 MeV. For the identification of galactic sources, a modelling of the diffuse galactic emission is essential. Such a modelling at this time does not yet exist at the required degree of accuracy. Therefore, a second COMPTEL source catalogue will be produced after a detailed and accurate modelling of the diffuse interstellar emission has become possible. 

\end{abstract}

\section{Introduction}
  
COMPTEL has demonstrated that the sky is rich in phenomena that can be studied at MeV energies. A variety of gamma-ray emitting objects are visible either in continuum or line emission. Among the continuum sources are spin-down pulsars, stellar black-hole candidates, supernova remnants, interstellar clouds, nuclei of active galaxies, gamma-ray bursters, and the Sun during solar flares. Line detections have been made in the light of the 1.809 MeV $^{26}$Al line, the 1.157 MeV $^{44}$Ti line, the 847 and 1238 keV $^{56}$Co lines, and the neutron capture line at 2.223 MeV.   

COMPTEL has also measured the diffuse interstellar and cosmic gamma radiation, whose properties are described elsewhere (\cite{strong96,strong99,bloemen99a,kappadath96,weidens99}). For the identification of galactic sources a modelling of the diffuse galactic emission is essential. Such a modelling at this time does not yet exist at the required degree of accuracy. 

This paper is restricted to all those sources that have been definitely or marginally detected so far, and provides upper limits to the MeV flux from different types of objects. A second COMPTEL source catalogue will be produced after a detailed and accurate modelling of the diffuse interstellar emission has become possible. 

The Compton Observatory was launched on April 5, 1991 by the Space Shuttle Atlantis. During Phase-I of the Compton Observatory Program, a full-sky survey - the first ever in gamma-ray astronomy - was performed. Phase-I ended on November 17, 1992. Observations during the subsequent phases of the program resulted in deeper exposures and complemented the survey. This source catalogue is mainly restricted to the results from the first five years of the mission (up to Phase IV/Cycle-5). In a few cases, more recent results have been added.


\section{Instrument description and data analysis}

COMPTEL was designed to operate in the energy range 0.75 to 30 MeV. It has a large field-of-view of about 1 steradian, and different sources within this field can be resolved, if they are more than about 3 to 5 degrees away from each other (energy dependent). The resulting source location accuracy is of the order of 1$^{\circ}$. The COMPTEL energy resolution of 5 \% to 10 \% FWHM is an important feature for gamma-ray line investigations. A detailed description of COMPTEL is given by \cite{schoenf93}.

COMPTEL consists of two detector assemblies, an upper one of liquid scinitillator NE213, and a lower one of NaI (Tl). A gamma-ray is detected by a Compton collision in the upper detector and a subsequent interaction in the lower detector. 

The arrival direction of a detected gamma ray is known to lie on a circle on the sky. The center   
of each circle is the direction of the scattered gamma-ray, and the radius of the   
circle is determined by the energy losses E1 and E2 in both interactions. The detected   
photons are binned in a 3-dimensional (3-D) data space, consisting of the scatter direction    
(defined by the two angular coordinates $\chi$ and $\psi$) and by the scatter angle   
$\phibar$ (derived from the measured energy losses in both interactions). Each detected photon is represented by a single point in the 3-D dataspace.   
The signature of a point source with celestial coordinates ($\chi_{o}$, $\psi_{o}$) is a cone of 90$^\circ$ opening angle with its axis parallel to the $\phibar$-axis. 
The apex of the cone is at ($\chi_{o}$,   
$\psi_{o}$). Imaging with   
COMPTEL involves recognizing the cone patterns in the 3-D dataspace. Two main techniques   
are applied: one is a maximum-entropy method that generates model-independent   
images (\cite{strong92}) and the other one is a maximum-likelihood method that is   
used to determine the statistical significance, flux and position uncertainty of a source (\cite{deboer92}).
 
Significances are derived in this method from the quantity -2 ln $\lambda$, where $\lambda$ is the maximum likelihood ratio L(B) / L(B+S); B represents the background model and S the source model (or sky intensity model). In a point-source search, -2 ln $\lambda$ formally obeys a $\chi_{3}^{2}$ distribution; in studies of a given source, $\chi_{1}^{2}$ applies. [This allows to translate a measured L value into a corresponding probability for it being a noise fluctuation, equivalent to a Gaussian $\sigma$ description of significances. In studies of a 'known' source, $\sigma$ = $\sqrt{-2 ln \lambda}$ applies.]

We verified by simulation that the shape of the probability density distribution for our application of the likelihood analysis to the COMPTEL data is indeed Gaussian. Furthermore, we calibrated by the same simulations the number of independent 'trials' we make in a search for source, taking into account the total sky area searched (see \cite{deboer92}).

Our threshold for detection is a chance probability of 99.7 $\%$, corresponding to a 3$\sigma$ Gaussian significance. The applicable statistics, i.e. the relevant trails, are discussed for each source in the original publication.

The sensitivity of COMPTEL is significantly determined by the instrumental background. A substantial suppression is achieved by the combination of effective charged-particle shield detectors, time-of-flight measurement techniques, pulse-shape discrimination, Earth-horizon angle cuts and proper event selections in energy and $\phibar$-space.  %

The application of the imaging techniques requires an accurate knowledge of the    instrumental and cosmic COMPTEL background. A variety of background models has   been investigated and is being used. In one method, the background is derived from averaging high-latitude observations.   
This assumes that the background has a constant shape in   
the instrumental system in at least the spatial coordinates (but not in Compton scatter angle) for all observations, and it also assumes that the extragalactic source   
contribution is small and smeared out by the averaging process (see \cite{strong99}). A second    
method derives the background from the data that are being studied itself.   
This is accomplished by applying a low-pass filter to the 3-D data, which   
smooths the photon distribution and eliminates (in the first   
approximation) the source signatures (e.g. \cite{bloemen94}). By applying iterations of this process the background estimate can be improved, further. All viewing periods have to be handled separately to account for changes of the background during the mission ({\cite{bloemen99a}). For line studies, we estimate the background below an underlying      
cosmic gamma-ray line by averaging the count rate from neighbouring energy   
intervals (\cite{diehl94}). 

For sources within the Galactic plane the global diffuse emission from the Galaxy is modelled by fitting a bremsstrahlung, and an inverse Compton component to the data. Also an isotropic component is added to these fits to represent the cosmic gamma-ray background. The amplitude of each of these components is obtained as free parameter from the fits. It has to be admitted, however (see above), that the modelling of the plane, at present, does not yet achieve the required degree of accuracy. 

In addition to the normal double-scatter mode of operation, two of the NaI crystals in the lower detector assembly of COMPTEL are also operated simultaneously as burst detectors. These two modules are used to measure the time history and energy spectra of cosmic gamma-ray bursts and solar flares. 

Hence, solar flares and cosmic gamma-ray bursts can be measured in the telescope mode (provided the event was within the field-of-view of the instrument), and in the burst mode.

In its telescope mode COMPTEL has an unprecedented sensitivity. Within a 2-week observation period it can detect sources, which are about 10-times weaker than the Crab. By adding up all data from a certain source that were obtained over the entire duration of the mission, higher sensitivities can be obtained. Table 1 summarizes the achieved point-source sensitivities for a 2-week exposure in Phase-I of the mission (t$_{\mbox{eff}}$ $\sim$ 3.5 $\cdot$ 10$^{5}$ sec), and for the ideal cases, when all data from a certain source in the Galactic center or anticenter (where the exposure is highest) are added from either Phase-I to III (t$_{\mbox{eff}}$ $\sim$ 2.5 $\cdot$ 10$^{6}$ sec) or Phase-I to IV/Cycle-5 (t$_{\mbox{eff}}$ $\sim$ 6  $\cdot$ 10$^{6}$ sec).
%
%
%
%

From this table rough upper limits can be derived for those objects, which are not contained in the later tables 10 to 12 by deriving the effective exposure from Fig.1.

%
%
%
%

\section {The first COMPTEL source catalogue}

This section consists of four different parts. The first part (Sect. 3.1) lists all observations on which the catalogue is based. Sect. 3.2 contains COMPTEL all-sky maps in continuum and line emission. Sect. 3.3 is the catalogue of detected sources, which is subdivided into detections of spin-down pulsars, galactic sources ($\mid$b$\mid$ $<$ 10$^{\circ}$), active galactic nuclei, unidentified high-latitude sources, gamma-ray line sources, gamma-ray burst sources within the COMPTEL field-of-view, and solar flare detections. Sect. 3.4 lists COMPTEL upper limits on source candidates, namely galactic objects, active galactic nuclei, and possible gamma-ray line sources.
 
\subsection{Observations and exposure maps}

This first COMPTEL source catalogue contains mainly results from Phase-I to IV/ Cycle-5 of the Compton mission. The relationship of the viewing periods (VP) to the actual dates of the observations is given in Table 2 (for completeness, the data of Phase-IV/Cycle-6 and 7 have been added in the table). The table also lists the pointing direction of the z-axis (COMPTEL telescope axis) in celestial coordinates, the duration of the pointing and the effective COMPTEL observation time. 

The effective COMPTEL exposure of the entire sky from the sum of all observations from the beginning of the mission to Phase-IV/Cycle-5 and Phase IV/Cycle-7 are illustrated in Fig. 1. The deepest exposures were obtained in the Galactic center and anticenter region, where effective observation times up to 6 $\cdot$ 10$^{6}$ seconds have been obtained (see also Table 1).

\subsection{COMPTEL all-sky maps}

COMPTEL all-sky maps exist for continuum emission in the three standard energy ranges 1--3 MeV, 3--10 MeV, and 10--30 MeV, and for the 1.8 MeV line from radioactive $^{26}$Al. These maps are shown in Fig. 2 and 3.

Fig. 2 is a maximum-entropy map using all data from Phase I to Phase IV/Cycle-6 (Strong et al., 1999). The background method used in this map is based on averaging high-latitude observations (see Sect. 2). 

%
%
%
%

Well known sources appear in the map:

Crab (l=184.5$^{\circ}$, b=5.9$^{\circ}$), Vela (263.6$^{\circ}$, -2.5$^{\circ}$) above 3 MeV, Cyg X-1 (71.1$^{\circ}$, +3.3$^{\circ}$), as well as striking excesses at (18$^{\circ}$,0$^{\circ}$) and near the Galactic center. At higher latitudes the sky is dominated by extragalactic sources: Cen A (309$^{\circ}$, +19$^{\circ}$) below 10 MeV (\cite{steinle95}), 3C 273 (290$^{\circ}$, +64$^{\circ}$) and 3C 279 (305$^{\circ}$, +57$^{\circ}$) (\cite{williams95b}). Various 'MeV blazars': 3C 454 (86$^{\circ}$, -38$^{\circ}$), PKS 0208-512 (276$^{\circ}$, -62$^{\circ}$), GRO J 0516-609 (270$^{\circ}$, -35$^{\circ}$) appear in one or more of the energy ranges. Next to the Crab the quasar PKS 0528+134 is clearly visible (\cite{collmar94}, \cite{collmar97a}). Details on these sources can be found in \cite{blom96} and 
references therein. 

Note that since these sources are variable, their appearance in these time-averaged maps may not reflect their published fluxes or spectra precisely. Another interesting feature is the apparent presence of significant areas of diffuse emission away from the plane; in particular, the regions around (170$^{\circ}$, +50$^{\circ}$) and (85$^{\circ}$, 35$^{\circ}$), which have been presented as candidates for assocations with high-velocity cloud complexes (\cite{blom97b}). Details in the structure of this emission should, however, be viewed with caution and are under further study. 

An alternative approach to derive all-sky continuum maps is described in \cite{bloemen99a}). This is a new approach combining model fitting, iterative background modelling and maximum entropy imaging, using the first five years of COMPTEL observations. On a coarse scale the maps derived by both methods are similar. However, on a fine scale, there are differences which are not yet fully understood. The uncertainties especially effect the identification of sources in the Galactic plane (see above).

%
%
%
%

Fig. 3 is a COMPTEL maximum-entropy map at 1.809 MeV from all observations up to VP 522.5 (\cite{oberlack97}). The brightest regions are in the inner Galaxy (-35$^{\circ}$ $<$ l $<$ +35$^{\circ}$), near Carina (l $\sim$ 285$^{\circ}$), and Vela (l $\sim$ 267$^{\circ}$). Other regions of enhanced emission are Cygnus (l $\sim$ 80$^{\circ}$), and Aquila (l $\sim$ 45$^{\circ}$). 

More recently, new 1.809 MeV COMPTEL all-sky maps have been produced using different imaging and background modelling methods. In one case (\cite{knoed99}) the analysis uses a multi-resolution version of the Richardson-Lucy algorithm, based on wavelets. In the other case (\cite{bloemen99b}), the maximum entropy method is combined with model fitting and iterative background modelling. All three maps are consistent with each other within their statistical and systematic uncertainties, although the multi-resolution map shows substantially less structure along the Galactic plane.  

So far, no all-sky map in the light of the 1.157 MeV $^{44}$Ti gamma-ray line exists. But imaging analysis along the Galactic plane have been performed by \cite{dupraz97} for data from Phase I to III, and by \cite{iyudin99} from Phase I to VI/Cycle-6. Two $^{44}$Ti gamma-ray line sources have been discovered so far, these are Cas A (\cite{iyudin97a}) and GRO J 0852-4642 (\cite{iyudin98}), a supernova remnant near the Vela region (\cite{aschenbach98}).  

%
%
%
%

A maximum-entropy map in the light of the 2.2 MeV neutron-capture line based on data from the first five years of the mission (VP 1.0 through VP 523.0) has been produced by \cite{mcconnell97b}. In general, the sky at 2.2. MeV is relatively featureless, e.g. the galactic plane is not visible. There is, however, evidence for significant ($\sim$ 3.7$\sigma$) emission from a point-like feature near (l, b) = (300.5$^{\circ}$, -29.6$^{\circ}$), the origin of which remains unknown at this time (\cite{mcconnell97b}). Flux limits for any candidate source are typically in the range (1 to 2.0) $\cdot$ 10$^{-5}$ cm$^{-2}$ sec$^{-1}$ (at the 3$\sigma$ significance level).

Fig. 4 is an all-sky map of the statistical location contours of 31 gamma-ray bursters, which happened to be in the COMPTEL field-of-view from the beginning of the mission up to viewing period 419.5 \cite{kippen98a}. 
%
%
%
%
%

\subsection{COMPTEL source detections}

The COMPTEL source detections are summarized in Table 3 to 9:  
   
\begin{itemize}  
\item Table 3:  Detected Spin-Down Pulsars 
\item Table 4:  Galactic Sources $\mid$b$\mid$ $<$ 10$^{\circ}$  
\item Table 5:  Active Galactic Nuclei
\item Table 6:  Unidentified High-Latitude Sources  
\item Table 7:  Gamma-Ray Line Sources
\item Table 8:  Gamma-Ray Burst Locations
\item Table 9:  Solar-Flare Detections.  
\end{itemize}
\medskip
\noindent

Out of the 7 spin-down pulsars listed in Table 4 COMPTEL has made firm detections from PSR B0531+21 (Crab), PSR B0833-45 (Vela), and from PSR B1509-58. The analysis of the Vela pulsar, however, is not yet finally settled; the results presented are presently based on Phase 0 and I, only. Because of the good statistics, the Crab pulsar fluxes are listed for smaller energy intervals than the standard ones (0.75--1, 1--3, 3--10 and 10--30 MeV). Only indications for emission in the COMPTEL energy range were found from the four pulsars PSR B1951+32, PSR J0633+1746 (Geminga), PSR 0656+14, and PSR B1055-52. 

The Galactic Sources with $\mid b \mid$ $<$ 10$^{\circ}$ listed in Table 4 are all objects, which were seen by at least one other experiment of the Compton Observatory. These are Cyg X-1, the two EGRET sources 2EG 2227+61 and 2EG J0241+6119 (which both are also COS-B sources), Nova Persei 1992 (GRO J0422+32), the Crab nebula, and an unidentified bright source at l = 18$^{\circ}$ within the plane (coincident with the EGRET source 2ES J1825-1307). The COMPTEL fluxes for these sources are listed in the standard energy ranges. For some of the sources the fluxes are also given for other energy intervals. The possible contribution of the diffuse galactic emission to the listed source fluxes constitutes a basic uncertainty (see column 11 of the Table). 

Nine of the Active Galactic nuclei listed in Table 5 are of the $\gamma$-ray Blazar type, discovered by EGRET (with the exception of 3C 273, earlier discovered by COS-B, \cite{swanenburg78}). The only non-blazar type object in the table is the radio galaxy Cen A. All $\gamma$-ray blazars are highly variable in intensity. 

Three of the five unidentified high-latitude sources in Table 6 are not point-like, but cover an extended region. Their extent may actually be due to a larger number of - so far - unresolved point sources (GRO J 1823-12 and the two High-Velocity Cloud complexes). 

The gamma-ray line sources listed in Table 7 are ordered with increasing line-energy. Apart from the four point-like sources SN 1991T, Cas A, Carina, and the SNR GRO J0852-4642, also the three extended emission regions of the inner Galaxy and of the Cygnus and Vela regions are included in the table. 

The 31 gamma-ray bursts listed in Table 8 were all recorded in the \lq telescope mode\rq\ . The error radius of the burst location (column 4) is defined as the angular radius, having the same area as the irregularily shaped COMPTEL 1$\sigma$ confidence region (see also Fig. 5). Columns 5, 6, and 7 provide informations on the COMPTEL accumulation time, the measured 0.75 to 30 MeV fluence and the COMPTEL detection significance of each burst. The parameters for a power-law fit to the spectrum of each burst are listed in columns 8 and 9. Column 10 states, whether variability of the burst spectrum during accumulation was observed in the more sensitive \lq burst mode\rq\ . 

Solar Flare gamma-ray measurements taken in the burst mode are summarized in Table 9. Column 1 contains the COMPTEL flare identification, including the year/month/day and UT of the flare start (taken from the GSFC Solar Data Analysis Center (SDAC)). Truncated Julian Day, in column 2 and column 3, contains the GOES X-ray start time (taken from the Solar Geophysical Reports). The X-ray classification is in column 4 and the BATSE flare number, as assigned at the SDAC at Goddard Space Flight Center, is in column 5. The corresponding BATSE trigger number is in column 2. The COMPTEL data measured in this burst mode (see Sect. 6) have been inspected and the flare (integration time) duration as judged by the visible signal in the spectrometer is listed in column 7 with the corresponding integrated counts in the full energy range in column 8. The peak counts (with an integration time of x s) are listed in column 9. The peak count rate as measured by BATSE is in column 10 as obtained from the SDAC at GSFC. Finally, in column 11 is the integrated amount of spacecraft and instrument material between the spectrometer and the Sun. This material attenuates the gamma-ray flux and degrades the spectrum. A small number is better. Large amounts of intervening material can, in principle, be modeled away when producing a photon spectrum, but the results have large errors and are often not unique.    

\subsection{COMPTEL upper limits to source candidates}

The limits are given at the 2$\sigma$ confidence level for the following objects:

\begin{itemize}
\item Table 10:  Galactic Objects $\mid$b$\mid$ $<$ 10$^{\circ}$  
\item Table 11:  Active Galactic Nuclei
\item Table 12:  Possible Sources of Gamma-Ray Line Emission  
\end{itemize} 

\medskip
\noindent

Table 10 contains upper limits to Galactic Sources with $\mid b \mid$ $<$ 10$^{\circ}$. This source list is restricted to black-hole candidates. Due to the above mentioned, still existing, uncertainty in modelling the diffuse Galactic emission, the source limits are at present rather conservative.  

Presented in Tables 11a and 11b are the
cumulative two-sigma upper limits to the MeV-emission measured with
COMPTEL from active galactic nuclei (AGN) and other unidentified gamma-ray
sources detected at high Galactic latitudes. These limits were derived
using composite COMPTEL all-sky maximum-likelihood maps for the 4.5 year
period covering Phases I through IV/Cycle-4 of the CGRO mission (1991- 1995). A description of the data-processing procedure used to obtain the
composite all-sky maps can be found in \cite{stacy97}. 

In the choice of candidate objects, emphasis was placed on known or suspected
gamma-ray sources, particularly those detected in neighbouring energy bands
to COMPTEL by the CGRO/EGRET and OSSE instruments (e.g., \cite{montigny95}, \cite{mcnaron95}). The flux-extraction routine applied
to the composite all-sky maps computes average output values within
one-pixel radius of a specified source location.  To minimize the number
of spurious false detections, only those objects for which the summed
log-likelihood ratio exceeds the equivalent of a three-sigma source
detection (adopting $\chi^{2}_{1}$ statistics, appropriate for a
previously known source at a specified location) are considered to exhibit
potentially significant MeV emission. 

Table 11a lists the COMPTEL cumulative upper limits for MeV-emission from
AGN through Phase IV/Cycle-4 of the CGRO mission. Column 1 gives the object name in
coordinate format; column 11 gives another common name for the source;
column 10 lists the object \lq type\rq\, which is either the object class (SY
for Seyfert galaxy, from the target list of \cite{maisack95}), or a
reference to a previously reported gamma-ray detection of this source (1EG
for the First EGRET Catalog of \cite{fichtel94}, 2EG for the Second
EGRET Catalog of \cite{thompson95}, 2EGS for the Supplement to the
Second EGRET Catalog of \cite{thompson96}). 

Table 11b lists the COMPTEL cumulative upper limits for MeV-emission from
unidentified high-latitude gamma-ray sources detected by the CGRO/EGRET
instrument. In both Tables 11a and 11b the recommended COMPTEL
team-standard corrections (for time-of-flight effects, livetime, etc.) are
applied to obtain final fluxes and upper limits (see \cite{diehl96}). 

Inspection of Tables 11a and 11b shows only in a few isolated cases the 
cumulative detection with COMPTEL of significant emission from high-latitude sources. This is no contradiction to the detections listed in Table 5. Note that all these sources are time-variable, and their detection in individual viewing periods does not mean that they are visible in cumulative maps. 

In general, the flux limits presented in Table 11a show that COMPTEL does not
detect cumulative MeV-emission from a majority of the extragalactic
blazar sources detected by EGRET. This result is similar to that obtained
by \cite{blom97}, for the case of individual CGRO viewing periods. 
Ultimately, these cumulative results will be further compared with other
source studies for individual CGRO viewing periods, and used in
statistical investigations of source properties by object class (e.g. 
\cite{blom97}, \cite{williams97}). 

The upper limits to possible sources of gamma-ray line emission in Table 12 are again ordered with increasing line energies. The line sources considered are SN 1993J ($^{56}$Ni $\rightarrow$ $^{56}$CO $\rightarrow$ $^{56}$Fe decay), four supernovae as possible $^{44}$Ti line emitters at 1.157 MeV, eleven recent novae as possible sources of $^{22}$Na line emission at 1.275 MeV, five nebula or cloud complexes as 1.809 MeV $^{26}$Al sources, and six possible 2.223 MeV neutron capture sources. 

\section{Conclusion}

COMPTEL has opened the MeV gamma-ray band (0.75 to 30 MeV) as a new window to astronomy. The data from five years of COMPTEL observations provided the first COMPTEL source catalogue. It is largely restricted to previously published results, and contains firm as well as marginal detections of continuum and line emitting sources, and presents upper limits for various types of objects. The number of most significant detections ($>$ 3$\sigma$) are 32 for steady sources and 31 for gamma-ray bursters (see Table 13). Six of the listed sources extend over larger areas. Their extent may actually be due to a larger number of so far unresolved point sources. This may be especially true for the Cygnus region in 1.809 MeV, for GRO J1823-12 and for the two HVC complexes. A second COMPTEL source catalogue will be produced after a more accurate modelling of the diffuse interstellar emission has become possible. 

\begin{acknowledgements} 
The COMPTEL project is supported by the German government through DLR grant 50 Q 9096 8, by NASA under contract NAS5-26645, and by the Netherlands Organization for Scientific Research NWO. The authors acknowledge the efforts of M. Chupp, H. Haber, J. Englhauser and R. Georgii in implementing the various tables. 
\end{acknowledgements}

\clearpage


\begin{figure*} [p]
\begin{center}  
\psfig{figure=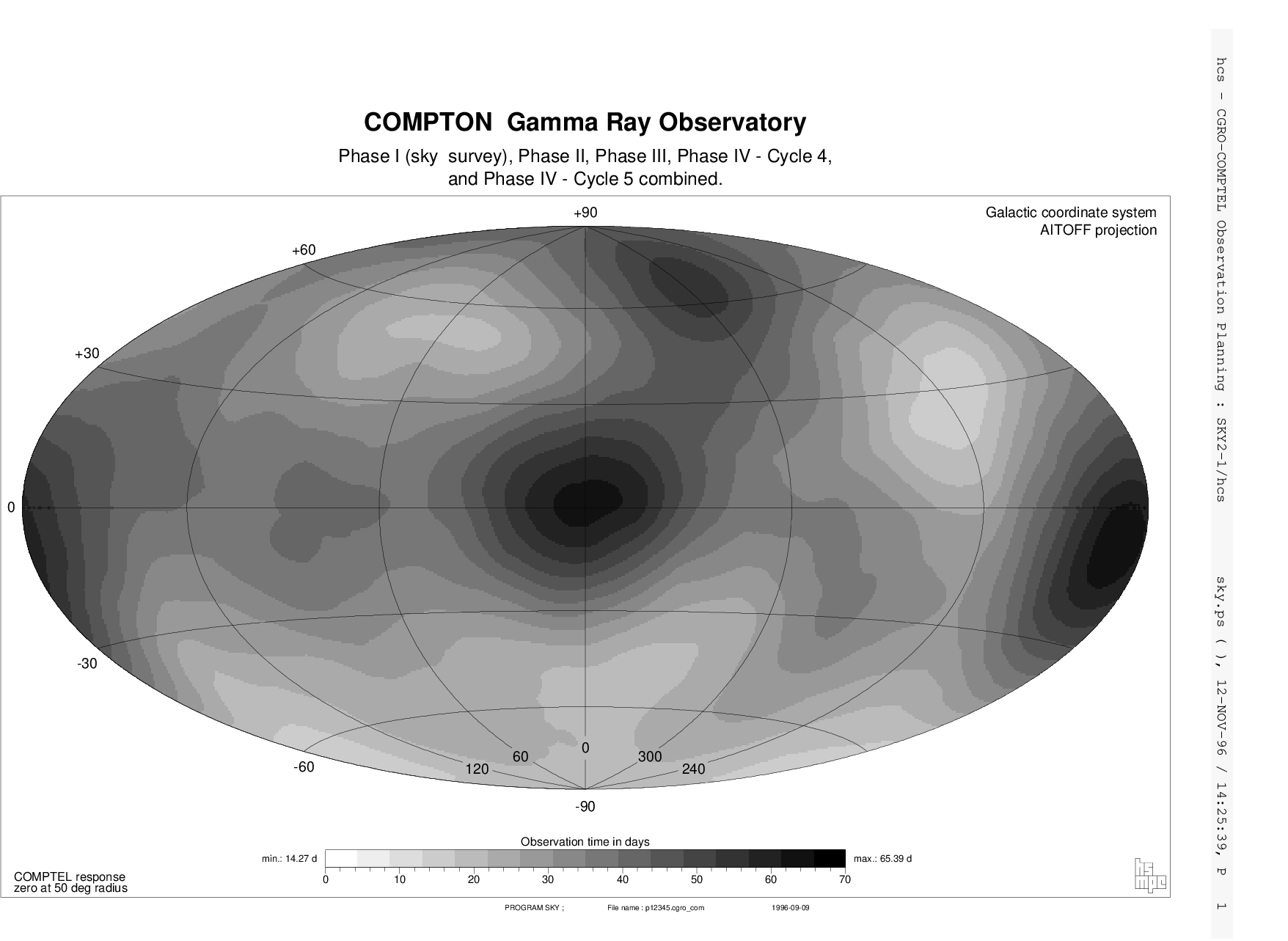,height=15cm,bbllx=0.0cm,bblly=0.0cm,bburx=20.1cm,bbury=28cm,angle=-90,clip=}
\vskip .2cm
\psfig{figure=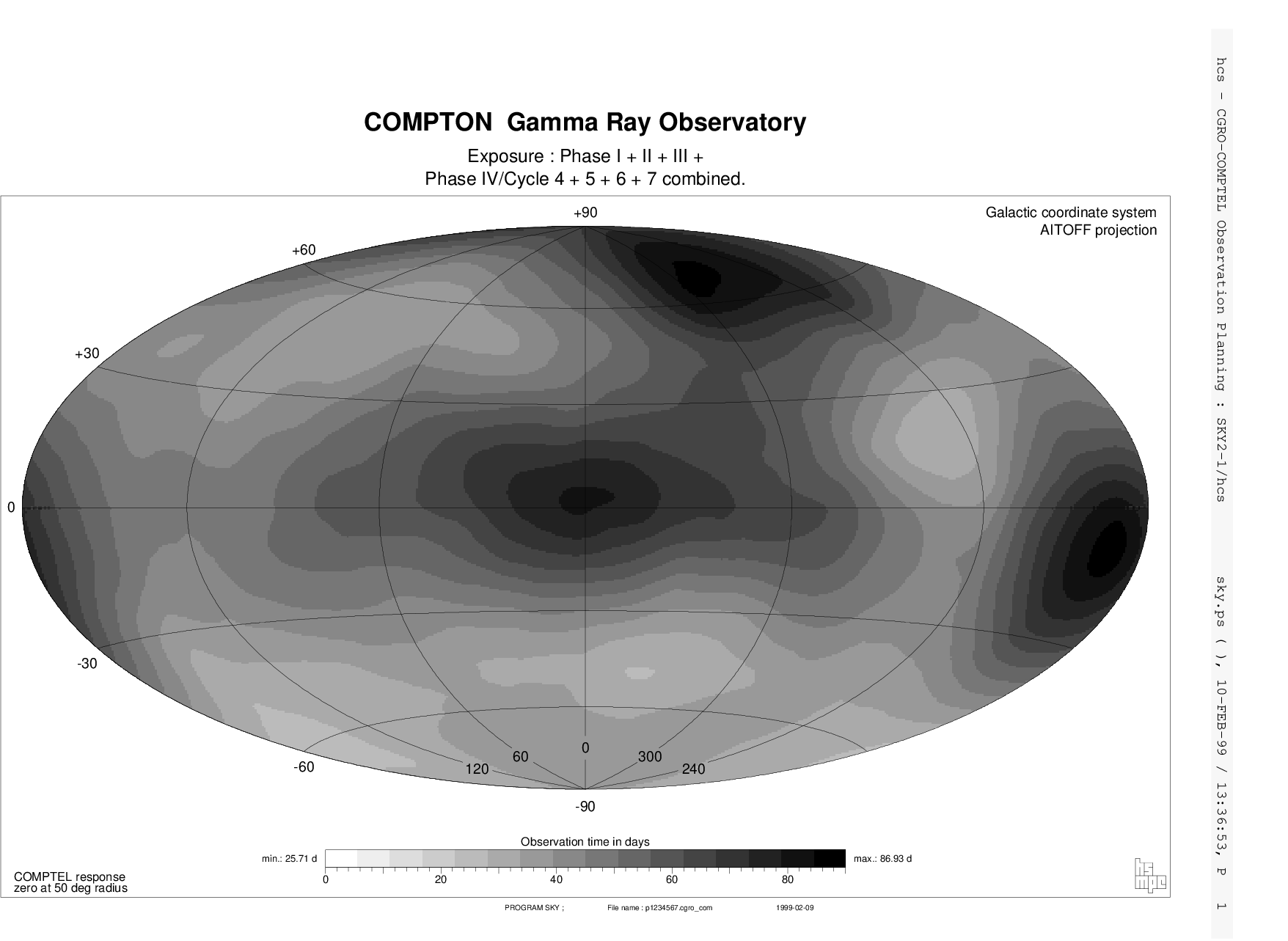,height=15cm,bbllx=0.0cm,bblly=0.0cm,bburx=20.1cm,bbury=28cm,angle=-90,clip=}
\caption{Effective exposure of COMPTEL from the sum of all observations. Top: Sum of all observations up to Phase IV/Cycle-5. Bottom: Sum of all observations up to Phase IV/Cycle-7.
}
\end{center}
\end{figure*}

\clearpage
 
%


\begin{figure*} [p]
\begin{center}
\large

1--3 MeV

\psfig{figure=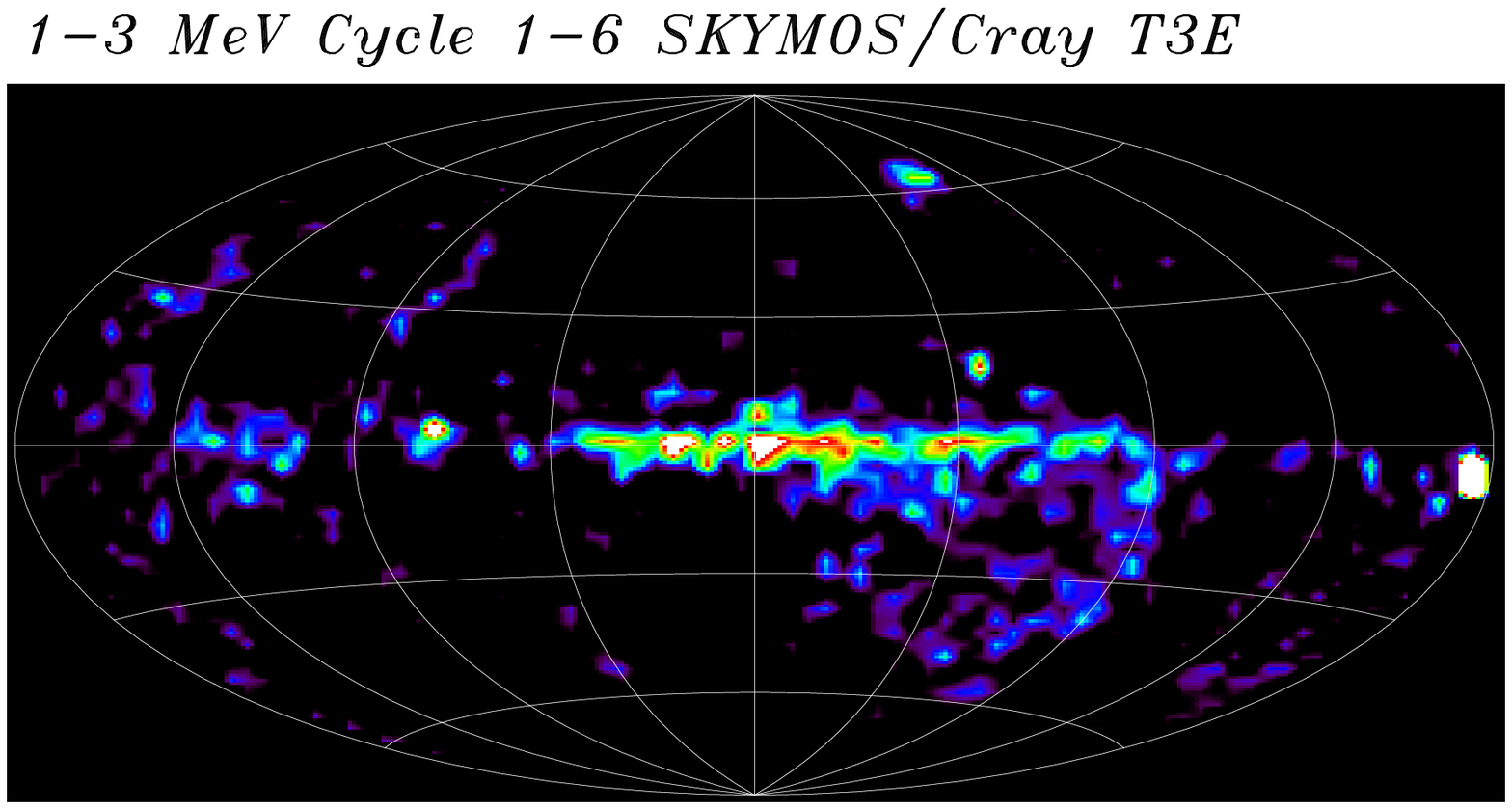,width=140mm,bbllx=60,bblly=145,bburx=563,bbury=390,clip=}

\vskip .1cm
3--10 MeV

\psfig{figure=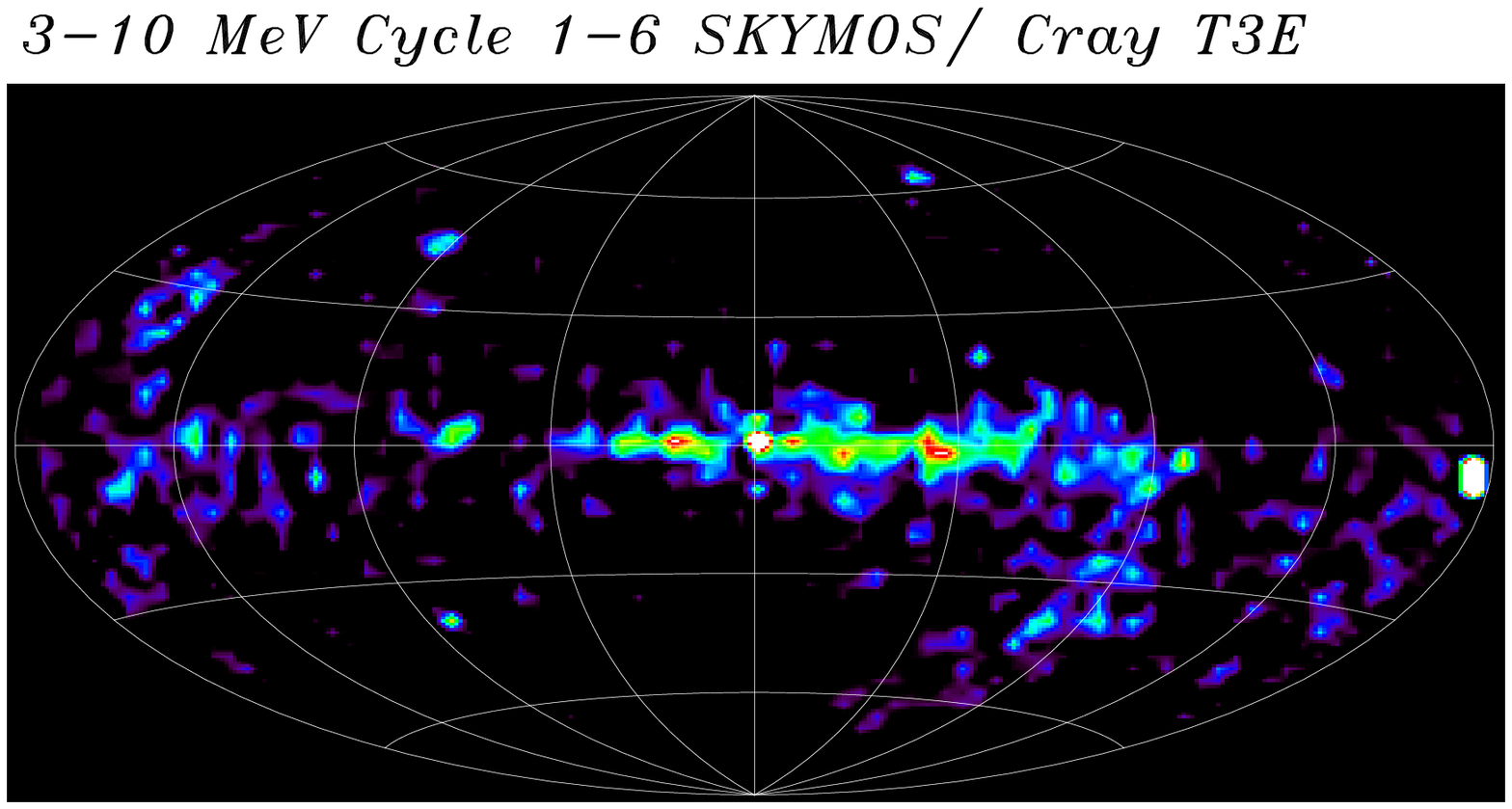,width=140mm,bbllx=60,bblly=145,bburx=563,bbury=390,clip=}

\vskip .1cm
10--30 MeV

\psfig{figure=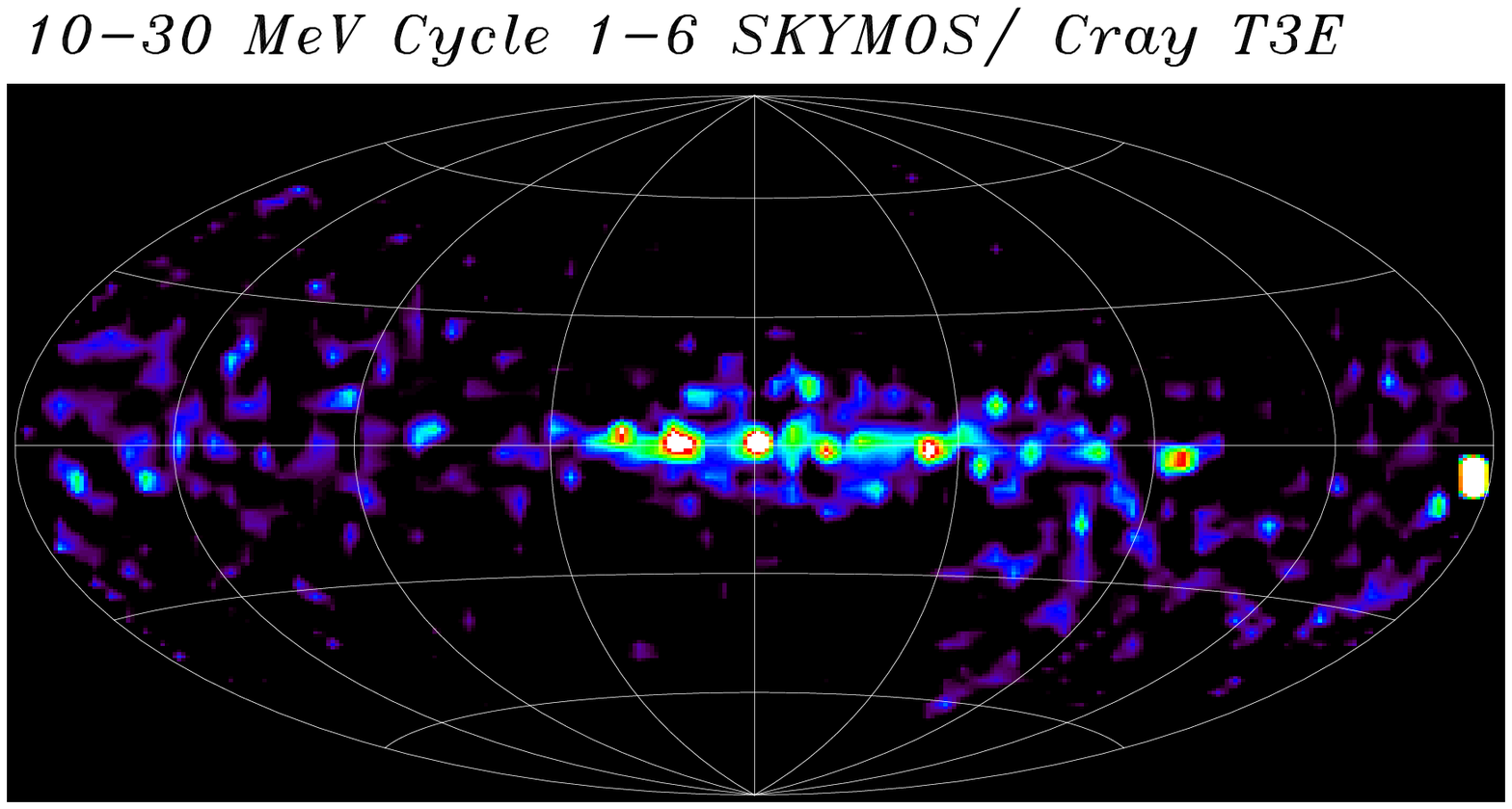,width=140mm,bbllx=60,bblly=145,bburx=563,bbury=390,clip=}

\caption[]{Full-sky maximum entropy intensity maps from all data between Phase I to IV/Cycle-6. Energy ranges are (from top to bottom): 1-3 MeV, 3-10 MeV, 10-30 MeV (from \cite{strong99}).
}
\end{center}
\end{figure*}

\clearpage


\begin{figure*} [p]
\psfig{figure=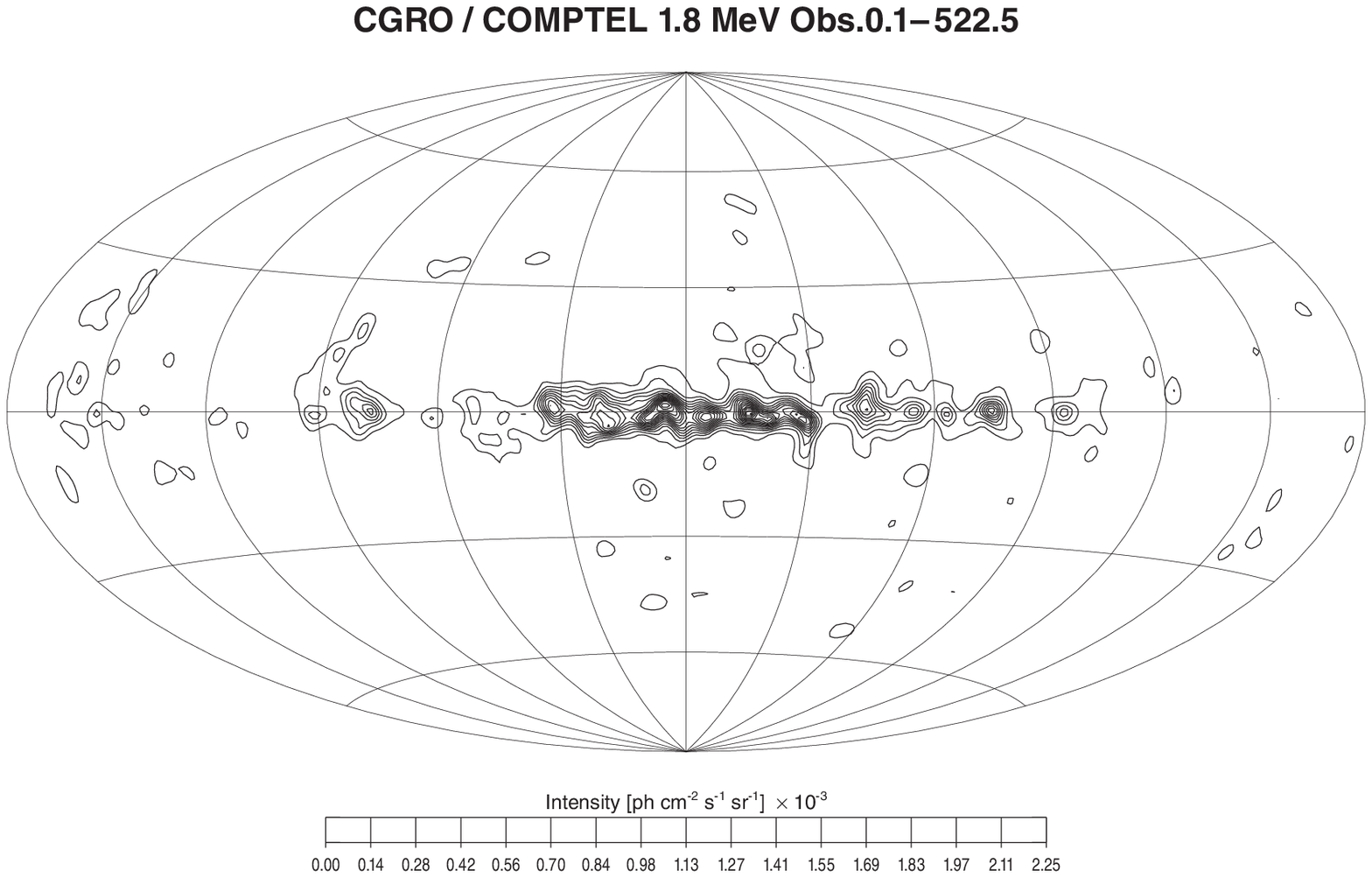,width=\textwidth}
\caption{COMPTEL maximum entropy map from VP 1 to VP 522.5 at 1.809 MeV (from \cite{oberlack97}).
}
\end{figure*}

\clearpage


\begin{figure*}[p]
\begin{center}
\psfig{figure=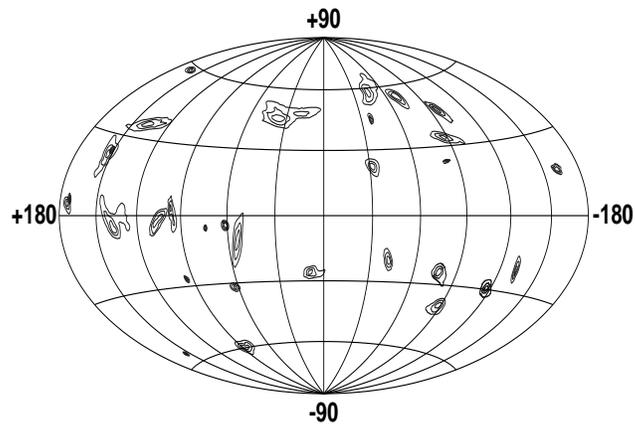,height=61mm,width=85mm}
\end{center}
\caption{Statistical (1-, 2-, and 3-sigma) location contours, in Galactic coordinates of 29 gamma-ray bursts observed through viewing period 419.5. The extent of the contours depends on the strength of the burst (from \cite{kippen98a}).}
\end{figure*}

\clearpage


\vspace{-5cm}
\begin{table*}
\begin{center}
\bf{ Table 1}: {COMPTEL 3$\sigma$ Point Source Sensitivity Limits}
\smallskip
\begin{tabular}{|c|c|c|c|}\hline
                      &  \multicolumn{3}{c|}{3$\sigma$ Flux Limits [10$^{-5}$ photons cm$^{-2}$ sec$^{-1}$]}  \\
 E$_{\gamma}[MeV]$    &  2 weeks in Phase 1 & Phase 1+2+3 & Phase 1+2+3+4 (Cycle-5)   \\
\hline
0.75 -- 1  &    20.1      &  7.4   &   3.7        \\
 1 -- 3    &    16.8      &  5.5   &   3.8         \\
 3 -- 10   &     7.3      &  2.8   &   1.7          \\
10 -- 30   &     2.8      &  1.0   &   0.8           \\
1.157      &     6.2      &  2.0   &   1.6            \\
1.809      &     6.6      &  2.2   &   1.6             \\
\hline
\end{tabular}
\caption[] {From this table rough upper limits can be derived for those objects, which are not contained in the later tables 10 to 12 by deriving the effective exposure from Fig. 1.
}
\end{center}
\end{table*}

\clearpage


\begin{table*}
{\bf Table 2}: {COMPTEL Observations, Pointings and Viewing Periods}

{\bf Phase I}
\begin{flushleft}
\begin{tabular}{||r|c|l|r|r|r|c||}
\noalign{\smallskip}
\hline
\hline 
&   &  & \multicolumn{2}{c|}{}  &  &   \\                                                              &   &  & \multicolumn{2}{c|}{}  &  &   eff. COMPTEL \\
 Viewing & start  & Z-axis & \multicolumn{2}{c|}{galactic}  & duration & observing time    \\
 Period  & (yy-mm-dd) & pointing   & LONG  & LAT   & (days) & (days)     \\  
&   &  &   &  &  & \\                                                      
\hline
\hline                                                                       
1.0    & 91-05-16 & Crab pulsar~~~& 190.9 & -4.7  & 14  & 5.3  \\
2.0    & 91-05-30 & Cyg X-1       & 73.3 &  2.6   &  9  & 3.8  \\
2.5    & 91-06-08 & Sun           & 194.9 & -7.3  &  7  & 3.5  \\
\hline                                                   
\end{tabular}
\end{flushleft}
\end{table*}

\clearpage

\footnotesize

\begin{table*}
{\bf Table 2}: {COMPTEL Observations, Pointings and Viewing Periods}

{\bf Phase II}
\begin{flushleft}
\begin{tabular}{||r|c|l|r|r|r|c||}
\noalign{\smallskip}
\hline
\hline                                                             
&   &  & \multicolumn{2}{c|}{}  &  &   \\                                                              &   &  & \multicolumn{2}{c|}{}   &  &    eff. COMPTEL \\
 Viewing & start  & Z-axis & \multicolumn{2}{c|}{galactic}  & duration & observing time    \\
 Period  & (yy-mm-dd) & pointing   & LONG  & LAT   & (days) & (days)     \\  
&   &  &   &  &  & \\                                                      
\hline
\hline
201.0   &  92-11-17 & Her X-1       &  66.8 &  39.3   &    7  &     1.1  \\ 
202.0   &  92-11-24 & Her X-1       &  70.9 &  40.5   &    7  &     1.2  \\ 
203.0   &  92-12-01 & Cygnus        &  77.9 &  0.7   &    7  &     2.0   \\
\hline                                                    
\end{tabular}
\end{flushleft}
\end{table*}

\clearpage

\footnotesize

\begin{table*}
{\bf Table 2}: {COMPTEL Observations, Pointings and Viewing Periods}

{\bf Phase III}
\begin{flushleft}
\begin{tabular}{||r|c|l|r|r|r|c||}
\noalign{\smallskip}
\hline
\hline                                                             
&   &  & \multicolumn{2}{c|}{}  &  &   \\                                                              &   &  & \multicolumn{2}{c|}{}   &  &    eff. COMPTEL \\
 Viewing & start  & Z-axis & \multicolumn{2}{c|}{galactic}  & duration & observing time    \\
 Period  & (yy-mm-dd) & pointing   & LONG  & LAT   & (days) & (days)     \\  
&   &  &   &  &  & \\                                                      
\hline
\hline
&   &   &   &  &  & \\
  303.7   &  93-10-17 & N Cyg 1992~~~~~~~~~&  89.1 &  7.8   &    2  &   0.5     \\
  304.0   &  93-10-19 & Virgo 278+67  & 278.2 &  66.7   &    6  &  1.7     \\
\hline                                                    
\end{tabular}
\end{flushleft}
\end{table*}

\clearpage

\footnotesize

\begin{table*}
{\bf Table 2}: {COMPTEL Observations, Pointings and Viewing Periods}

{\bf  Phase IV/Cycle-4}
\begin{flushleft}
\begin{tabular}{||r|c|l|r|r|r|c||}
\noalign{\smallskip}
\hline
\hline                                                             
&   &  & \multicolumn{2}{c|}{}  &  &   \\                                                              &   &  & \multicolumn{2}{c|}{}   &  &    eff. COMPTEL \\
 Viewing & start  & Z-axis & \multicolumn{2}{c|}{galactic}  & duration & observing time    \\
 Period  & (yy-mm-dd) & pointing   & LONG  & LAT   & (days) & (days)     \\  
&   &  &   &  &  & \\                                                      
\hline
\hline
&   &   &   &  &  & \\
  403.0   &  94-11-01 & Her X-1       &  58.2 &  37.5   &    8  &     2.1   \\
  403.5   &  94-11-09 & OJ 287        & 206.8 &  35.8   &    6  &     1.6   \\
\hline                                                    
\end{tabular}
\end{flushleft}
\end{table*}

\clearpage

\footnotesize

\begin{table*}
{\bf Table 2}: {COMPTEL Observations, Pointings and Viewing Periods}

{\bf  Phase IV/Cycle-5}
\begin{flushleft}
\begin{tabular}{||r|c|l|r|r|r|c||}
\noalign{\smallskip}
\hline
\hline                                                             
&   &  & \multicolumn{2}{c|}{}  &  &   \\                                                              &   &  & \multicolumn{2}{c|}{}   &  &    eff. COMPTEL \\
 Viewing & start  & Z-axis & \multicolumn{2}{c|}{galactic}  & duration & observing time    \\
 Period  & (yy-mm-dd) & pointing   & LONG  & LAT   & (days) & (days)     \\  
&   &  &   &  &  & \\                                
\hline
\hline
&   &   &   &  &  & \\
  517.0   &  96-03-05 & PKS 0208-512  & 276.8 & -59.6   &    13  &     3.5 \\
  516.1   &  96-03-18 & GRO J1655-40  & 341.1 &  5.6   &    3  &     0.8  \\
\hline                                                    
\end{tabular}
\end{flushleft}
\end{table*}

\clearpage

\footnotesize

\begin{table*}
{\bf Table 2}: {COMPTEL Observations, Pointings and Viewing Periods}

{\bf  Phase IV/Cycle-6}
\begin{flushleft}
\begin{tabular}{||r|c|l|r|r|r|c||}
\noalign{\smallskip}
\hline
\hline                                                             
&   &  & \multicolumn{2}{c|}{}  &  &   \\                                                              &   &  & \multicolumn{2}{c|}{}   &  &    eff. COMPTEL \\
 Viewing & start  & Z-axis & \multicolumn{2}{c|}{galactic}  & duration & observing time    \\
 Period  & (yy-mm-dd) & pointing   & LONG  & LAT   & (days) & (days)     \\  
&   &  &   &  &  & \\                                                       
\hline
\hline
&   &   &   &  &  & \\
  620.0   &  97-06-10 & GAL 16+4~~~~~~~~~&  15.8 &   3.6  &    7  &    2.4   \\
  621.5   &  97-06-17 & 3C 279        & 301.7 &  57.8  &    7  &    2.5      \\
\hline                                                    
\end{tabular}
\end{flushleft}
\end{table*}

\clearpage

\footnotesize
\begin{table*} [t]
\vspace{-8cm}
{\bf Table 2}: {COMPTEL Observations, Pointings and Viewing Periods}
{\bf  Phase IV/Cycle-7}
\begin{flushleft}
\begin{tabular}{||r|c|l|r|r|r|c||}
\noalign{\smallskip}
\hline
\hline                                                             
&   &  & \multicolumn{2}{c|}{}  &  &   \\                                                              &   &  & \multicolumn{2}{c|}{}   &  &    eff. COMPTEL \\
 Viewing & start  & Z-axis & \multicolumn{2}{c|}{galactic}  & duration & observing time    \\
 Period  & (yy-mm-dd) & pointing   & LONG  & LAT   & (days) & (days)     \\  
&   &  &   &  &  & \\                                                    
\hline
\hline
&   &   &   &  &  & \\
  730.6   &  98-08-26 & SN 1998bu     & 254.9 &   45.5  &   14  &     4.6   \\
  731.5   &  98-09-08 & SN 1998bu     & 235.7 &   55.4  &    7  &     2.2   \\
  732.5   &  98-09-15 & GAL 058-12    & 232.3 &   58.7  &    7  &     2.5    \\
  728.7   &  98-09-22 & J0218+4232    & 139.4 &  -18.7  &    3  &     1.1    \\
  729.5   &  98-09-25 & J1550-564     & 324.9 &  - 0.2  &   11  &     3.0    \\
  734.0   &  98-10-06 & GAL 190-70    & 204.1 &  -72.0  &   14  &     4.5   \\
  728.9   &  98-10-13 & J0218+4232    & 139.4 &  -18.7  &   21  &     6.1    \\
  734.5   &  98-11-03 & GAL 190-70    & 204.1 &  -72.0  &   14  &     4.6     \\
  736.0   &  98-11-10 & Orion B1      & 227.9 &  - 4.3  &    6  &     1.9    \\
  736.5   &  98-11-16 & Leonid Save   &  17.3 &  -37.5  &    2  &     ~~~     \\
  736.7   &  98-11-18 & Orion B1      & 227.9 &  - 4.3  &    6  &     1.9     \\
  737.0   &  98-11-24 & GAL 044-09    &  44.3 &  - 9.1  &    7  &     2.6    \\
          &  98-12-01 & End of  &     &         &       &            \\
          &           & Phase IV/Cycle-7~~~~~~~~~&       &         &       &          \\
&   &   &   &  &  & \\
\hline
\hline                                                    
\end{tabular}
\end{flushleft}
\end{table*}

\clearpage


\landscape
\footnotesize

{\bf Table 3}: {Spin-Down Pulsars}
\begin{flushleft}
\begin{tabular}{||l|c|l|l|c|c|c|c|c|c|c|l|c||}
\noalign{\smallskip}
\hline
\hline
 & & & & & &\multicolumn{5}{c|}{ }&  & \\
Pulsar & l & b &\ \ P & \ \ \.{P} & d 
       & \multicolumn{5}{c|}{Phase Averaged Flux (10$^{-5}$ cm$^{-2}$   
          s$^{-1}$)} & 
       VP or Phase & Ref. \\

 & & & & & & (0.75--1) & (1--3) & (3--10) & (10--30)  & Other energy  &  & \\
& [deg.] & [deg.] & \  [ms] & [$10^{-15}$] & \ \  [kpc] & [MeV] & [MeV] 
& [MeV] & [MeV] & ranges [MeV] & & \\

& & & & & & & & & & & & \\
\hline 
\hline
& & & & & &\multicolumn{4}{c|} {}& & &   \\
PSR B1951+32 & 68.77 & 2.82 & 39.53 & 5.849 & 2.5 
      & \multicolumn{4}{c|}{(2.3 $\pm$ 1.4)}  &    & Phase &  6   \\
 & & & & & & & & & & & I--IV/Cycle-6 &  \\
\hline
\hline
\end{tabular}
\end{flushleft}

\vskip6pt
\noindent
{\bf Notes}
 \vskip6pt
\noindent
(a)  Flux main pulse from timing analysis.
\noindent
(b)  Flux main pulse above background from spatial analysis.
\noindent
(c)  Total flux from spatial analysis.
\noindent
(d)  Note that the flux given in ref. (8) should be in units ph cm$^{-2}$ sec$^{-1}$ MeV$^{-1}$.

\vskip6pt
\noindent
{\bf References}
\vskip6pt
\noindent
(1)  \cite{much95a}.

\noindent
(2)  \cite{much97}.

\noindent 
(3)  \cite{kuiper96}.

\noindent  
(4)  \cite{hermsen97}. 

\noindent
(5)  \cite{schoenf95}.

\noindent
(6)  \cite{kuiper98a}.

\noindent
(7)  \cite{kuiper98b}. 

\noindent 
(8)  \cite{thompson99}.

\noindent
(9)  \cite{kuiper99a}.

\noindent
(10) \cite{kuiper99b}.   
                
\endlandscape

\clearpage


\landscape
\def \p{{$\pm$}}
\def\s{{$\sigma$}}
\footnotesize
{\bf Table 4}: {Galactic Sources $|b|< 10^\circ$}
\begin{flushleft}
\begin{tabular}{||l|r|r|c|c|c|c|c|c|c|c|c||}
\noalign{\smallskip}
\hline
\hline
 &  \multicolumn{2}{r|}{} &  \multicolumn{5}{c|}{} & & & & \\
Source  & \multicolumn{2}{r|}{Position} & \multicolumn{5}{c|}{Fluxes$^{\rm a}$
 (10$^{-5}$ photon cm$^{-2}$   
          s$^{-1}$)} & Spect. & VP or & Notes & Ref. \\ 
& l & b & (0.75--1) & (1--3) & (3--10) & (10--30)
   & Other Energy & Fits & Phase &  & \\
& & & & & & & Ranges & & & & \\
& [deg] & [deg] & [MeV] & [MeV] & [MeV]  & [MeV] &  [MeV]  & & & &  \\
& & & & & & & & & & &  \\
\hline
\hline
& & & & & & & & & & &  \\
GRO J 1823-12&18.5&-0.5& 4.1\p 1.7 & 9.9\p 1.5 & 3.5\p 0.6 & 1.0\p 0.2 & ~~~~~~~& f &  VP 1 to 522.5 & c & (9) \\
(2 EG J 1825-1307) & & & & & & & & & & &  \\
& & & & & & & & & & &  \\
\hline
& & & & & & & & & & &  \\
Cygnus X--1 & 71.3 & 3.1 & & & & $<$1.39 & 
 11.9\p 3.0 (0.75--0.85) &  
            & Phase I to III & b  & (3) \\
& & & & & & & 9.5\p 2.2 ((0.85--1) &  a & (up  to VP 318.1) & & (3a)\\
& & & & & & & 16.7\p 2.7 (1--2) & & & & \\
& & & & & & & 6.9\p 1.6 (3--5) & & & &\\
& & & & & & & $<$1.15 (5--10) & & & & \\
& & & & & & & & & & & \\
\hline
& & & & & & & & & & & \\
GRO J 2227+61 & 106.6 & 3.1 & 12.6\p 4.5 & 30.8\p 5.5 & $\leq$6.3 
& $\leq$1.48 & & e & VP  2+7+34 & & (6) \\
 (2 EG 2227+61) & & & & & & & & & & &  \\
 (2CG 106+1.5)  & & & & & & & & & & &  \\
& & & & & & & & & & &  \\
\hline
& & & & & & & & & & &  \\
GT 0236+610  & 135.7 & 1.1 & $<$ 4.8 & 11.2\p 3.2 & 5.0\p 1.2 
                      & 1.24\p 0.38 &
            & b & VP 15+31+34+211 & b &  (5) \\
(2EG J0241+6119) & & & $<$ 4.8 & 5.6 \p 3.0 & 3.2\p 1.3 & $<$ 1.3 
&  &  &  +319+325 & c &  \\ 
(2CG 135+01)  & & & & & & & & & & &  \\
 & & & & & & & & & & &   \\
\hline
 & & & & & & & & & & &   \\
Nova Per 1992  & 165.9 & -11.9 & $<$ 35.0 & & $<$11.9 & $<$2.8 
               & 28.0\p 10.0 (1--2) & c & VP 36.5 & b & (4) \\
(GRO J0422+32)  & &  & & & & & $<$8.1 (2--3) & & & & \\
 & & & 27.5\p 10.00 & & $<$6.0 & $<$2.2  & $<$15.0 (1--2)& & VP 39 & &  \\
 & & & & & & &  $<$6.6 (2--3) & & & & \\
 & & & & & & & & & & &  \\
\hline
& & & & & & & & & & & \\
Crab Unpulsed~~~~~~~ & 184.6 & -5.8 &   
    & & & &    34.07\p 1.57(0.78-0.96)  & & Phase I to IV/Cycle-5 & & \\
& & & & & & &  31.11\p 1.30(0.96-1.16)  & & & a,b  & ~~ \\
& & & & & & &  21.54\p 1.09(1.16-1.38)  & & &      &   \\
& & & & & & &  18.68\p 1.04(1.38-1.62)  & & &      &    \\
& & & & & & &  13.03\p 0.82(1.62-1.88)  & & &      &     \\
& & & & & & &   8.90\p 0.71(1.88-2.16)  & & &      &      \\
& & & & & & &   7.84\p 0.70(2.16-2.48)  & & &      &       \\
& & & & & & & & & & & \\
\hline\multicolumn{12}{r}{(Table 4 cont.)}\\
\end{tabular}
\end{flushleft}

\clearpage

\hoffset=-3cm
{\bf Table 4}: {Galactic Sources $|b|< 10^\circ$ (cont.)}
\begin{flushleft}
\begin{tabular}{||l|r|r|c|c|c|c|c|c|c|c|c||}
\noalign{\smallskip}
\hline
\hline
 &  \multicolumn{2}{r|}{} &  \multicolumn{5}{c|}{} & & & &  \\
Source  & \multicolumn{2}{r|}{Position} & \multicolumn{5}{c|}{Fluxes$^{\rm a}$
 (10$^{-5}$ photon cm$^{-2}$   
          s$^{-1}$)} & Spect. & VP or & Notes & Ref. \\ 
& l & b & (0.75--1) & (1--3) & (3--10) & (10--30)
   & Other Energy & Fits & Phase &  & \\
& & & & & & & Ranges & & & & \\
 & [deg] & [deg] & [MeV] & [MeV] & [MeV]  & [MeV] &  [MeV]  
 & & & &  \\
& & & & & & & & & & &  \\
\hline
\hline
& & & & & & & & & & & \\
Crab Unpulsed cont. 
& ~~&~~ & ~~~~~~~~~~~~~& ~~~~~~~~& ~~~~~~~~~& &   8.66\p 0.57(2.48-2.84)  & &~~~~~~~~~~~~~~~~~~~~~~~ &      &   \\
& & & & & & &   6.14\p 0.52(2.84-3.22)  & & &      & (1),(2) \\
& & & & & & &   4.41\p 0.46(3.22-3.62)  & & &      & (7)      \\
& & & & & & &   4.28\p 0.41(3.62-4.08)  & & &      &           \\
& & & & & & &   4.44\p 0.36(4.08-4.56)  & & &      &            \\
& & & & & & &   2.94\p 0.33(4.56-5.08)  & & &      &             \\
& & & & & & &   3.42\p 0.32(5.08-5.66)  & & &      &        \\
& & & & & & &   2.30\p 0.28(5.66-6.26)  & & &      &         \\
& & & & & & &   2.15\p 0.28(6.26-6.94)  & & &      &          \\
& & & & & & &   1.60\p 0.25(6.94-7.64)  & & &      &           \\
& & & & & & &   1.17\p 0.20(7.64-8.42)  & & &      &            \\
& & & & & & &   1.36\p 0.17(8.42-9.26)  & & &      &             \\
& & & & & & &   1.24\p 0.15( 9.26-10.16)& & &      &              \\
& & & & & & &   1.05\p 0.15(10.00-11.20)& & &      &               \\
& & & & & & &   1.12\p 0.14(11.20-12.48)& & &      &                \\
& & & & & & &   1.12\p 0.14(12.48-13.92)& & &      &          \\
& & & & & & &   0.62\p 0.13(13.92-15.52)& & &      &           \\
& & & & & & &   0.83\p 0.14(15.52-17.28)& & &      &            \\
& & & & & & &   0.50\p 0.14(17.28-19.28)& & &      &             \\
& & & & & & &   0.31\p 0.14(19.28-21.60)& & &      &              \\
& & & & & & &   0.23\p 0.15(21.60-24.08)& & &      &               \\
& & & & & & &   0.16\p 0.18(24.08-26.88)& & &      &                \\
& & & & & & &   0.13\p 0.27(26.88-30.00)& & &      &                 \\
& & & & & & & & & & & \\
\hline
& & & & & & & & & & & \\
Vela/Carina & 273$^{\circ}$~ & -6$^{\circ}$ & $<$ 6.0~~~~~~  & $<$ 5.4~~~ & 7.5\p 1.1 &
$<$ 1.1 &        &   &  VP 1 to 522.5~~~~~~~~~ & a,c,d & (8) \\
& & & & & & & & & & &  \\
\hline\multicolumn{12}{r}{(Table 4 cont.)}\\
\end{tabular}
\end{flushleft}

\clearpage

{\bf Spectral Fits}
\vskip6pt

a.  Wien spectrum with kT $\sim$ 190 keV; for fits to 1991 data only
    see McConnell et al. 1994.

b.  1.75 $\cdot$ 10$^{-4}$(E/1 MeV$^{-1.95}$ photon cm$^{-2}$ s$^{-1}$ 
    MeV$^{-1}$ (ref. 5).

c.  See ref. 4.

d.  (1.12\p 0.03) $\cdot$ 10$^{-4}$(E/3.5 MeV)$^{-(2.02\mbox{\small \p}0.03)}$ photon cm$^{-2}$  s$^{-1}$  MeV$^{-1}$ $<$ 10 MeV (ref. 7).

e.  Power-law spectrum $\sim$ E$^{-\alpha}$,  with $\alpha$ = 2.20 $\pm$ 0.14.

f.  Power-law spectrum $\sim$ E$^{-\alpha}$,  with $\alpha$ = 2.0.

\vskip6pt
{\bf Notes}
\vskip6pt

a.  Flux values with statistical 1 \s \ uncertainties and 2  \s \ upper limits
 are given.

b.  No model for Galactic diffuse emission is included in this analysis.

c.  Model for Galactic diffuse emission is included.

d.  The excess is extended and can be explained by a 2-source model with sources at (273.8, -3.5) and (271.9, -9.0).

\vskip6pt
{\bf References}
\vskip6pt
(1) \cite{much95a}.

(2) \cite{much95b}.

(3) \cite{mcconnell97b}.

(3a) \cite{mcconnell94}.

(4) \cite{vandijk95}.

(5) \cite{vandijk96}.

(6) \cite{iyudin97b}.

(7) \cite{meulen98}.

(8) \cite{meulen99}.

(9) \cite{bloemen00}.
 

\endlandscape

\clearpage


\footnotesize
{\bf Table 5}: {Active Galactic Nuclei}
 \vskip12pt
\begin{tabular}{||l|c|c|c|c|c|c|c|c|c||}
\hline\hline \noalign{\smallskip}
 & & \multicolumn{2}{c|}{} &  \multicolumn{4}{c|}{} & & \\
Source & Z-Value & \multicolumn{2}{c|}{Position} & \multicolumn{4}{c|}{Flux (10$^{-5}$ photon cm$^{-2}$ s$^{-1}$)}  & VP  & Ref. \\
 & & & & & & & & or &    \\
& & l & b & (0.75--1) & (1--3) & ~~(3--10)~~ & (10--30) & Phase &    \\
&  & [deg] & [deg] & [MeV] & [MeV] & [MeV]  & [MeV] & &  \\
& & & & & & & & &   \\
\hline
\hline
& & & & & & & & &    \\
CTA 102 & 1.037 & 77.4 & -38.6 & $<$10.4 & $<$12.1
 &  $<$7.4  & 1.4$\pm$1.1  &  VP 19   & (1) \\
(PKS 2230+114)& & & & $<$15.1 & $<$24.7 & $<$10.7 & $<$6.9 & VP  26+28  & \\ 
 & & & & $<$17.9 &  $<$30.0 & $<$13.5 & $<$6.5 & VP 37 & \\
 & & & & $<$7.8 &  $<$12.2 & $<$6.4 & 1.9$\pm$0.9 & Phase 1 & \\
& & & & & & & & &   \\
\hline
& & & & & & & & &   \\
3C 454.3 & 0.859 & 86.1 & -38.2 & $<$10.5 & $<$12.9 & 6.9$\pm$3.3 
             & $<$3.8 & VP 19 & (1) \\
(PKS 2251+15) & & & & $<$17.5 &  $<$23.0 & $<$8.1 & 6.1$\pm$1.80 & VP 26+28 & \\
 & & & & $<$20.2  & 12.4 $\pm$10.4 &  $<$13.0 & 2.2$\pm$1.8  & VP 37  & \\
 & & & & $<$10.4  &  $<$9.2  & 2.8$\pm$2.3 & 2.9$\pm$1.0  & Phase 1   &  \\
& & & & & & & & &   \\
\hline
& & & & & & & & &   \\
PKS 0528+134 & 2.06 & 191.4 & -11 & 9.4$\pm$8.2 & 9.7$\pm$6.3 & 4.4$\pm$2.5 
         & 3.2$\pm$1.0 & VP 0  & \\
(OG 147) & & & & 16.6$\pm$8.7  & $<$13.8 &  $<$7.7 & 3.0$\pm$1.0& VP 1  & \\
 & & & & $<$16.2  &  $<$21.8 & $<$9.4 & $<$3.2 & VP 2.5 & \\
 & & & & $<$15.6  &  $<$18.7 & $<$8.4 & $<$2.8 & VP 36/39  & \\
 & & & & $<$32.1 &  15.3 $\pm$12.3 &  $<$15.2$\pm$4.7 & 3.3$\pm$1.7  
                 & VP 213  & \\
 & & & & $<$24.4  &  $<$14.0$\pm$9.5 & $<$4.4$\pm$3.6  & $<$2.7 & VP 221  & \\
& & & & $<$20.9  &  $<$17.4 & 5.0$\pm$3.6 & $<$4.3 & VP 310  & \\
 & & & & 11.3$\pm$10.0 & 18.1 $\pm$8.3 & $<$6.3 & $<$3.1 & VP 321 
  &  (2) \\ 
 & & & & 23.8$\pm$7.9 & 20.6 $\pm$6.6  &  $<$5.0  & $<$1.8 & VP 337 
  & \\ 
& & & & $<$32.1 & 15.3$\pm$12.3 & 15.2 $\pm$4.7 & 3.3$\pm$1.7 &  VP213 
  & \\
 & & & & $<$8.7 & 5.8$\pm$3.4  & 4.3$\pm$1.4 & 2.0$\pm$0.5 & Phase I+II
 &  \\
 & & & & 4.2$\pm$3.4 & 8.5$\pm$2.7  & 2.3$\pm$1.0 & 1.3$\pm$0.4 
       & Phase I--III &  \\
 & & & &  & & & & &  \\
\hline
& & & &  & & & & &  \\
GRO J0516--609  & & 270 & -35 & $<$12.8 & $<$18.1 & $<$8.3 & $<$2.1 & VP 6 
  & (3) \\
(PKS 0506-612/& & & & $<$18.6 & 20.0$\pm$7.2 & $<$12.4 & $<$4.2 & VP 10
  & \\
PKS 0522-611) & & & & $<$11.0 & 9.2 $\pm$6.2 & 8.9$\pm$3.2 & $<$1.7 & VP 17 
  &  \\
& & & & $<$9.4  & 15.1$\pm$4.6 & 5.7$\pm$2.5 & $<$1.5 & VP 10+17   & \\
& & & & & & & & &     \\
\hline
& & & &  & & & & &  \\
PKS 0208-512 & 1.003 & 276.1 & -61.8 & & 21$\pm$9 & & & VP 6  & \\
(RX J02107--5100)& & & &  & $<$26 & & & VP 9  & \\
& & & &  & $<$15 & & & VP 10 &  \\
& & & &  & $<$18 & & & VP 13.5  & \\
& & & &  & $<$23 & & & VP 17  & (4) \\
& & & &  & 35$\pm$12 & & & VP 220  & \\
& & & &  & 44$\pm$9  & & & VP 224   & \\
& & & & $<$12 & 41$\pm$7 & $<$7 & $<$3 & Phase II  & \\
& & & &  & & & & &   \\
\hline\multicolumn{10}{r}{(Table 5 cont.)}\\
\end{tabular}

\clearpage

\begin{tabular}{||l|c|c|c|c|c|c|c|c|c||}
\hline\hline \noalign{\smallskip}
 & & \multicolumn{2}{c|}{} &  \multicolumn{4}{c|}{} & & \\
Source & Z-Value & \multicolumn{2}{c|}{Position} & \multicolumn{4}{c|}{Flux (10$^{-5}$ photon cm$^{-2}$ s$^{-1}$)}  & VP  & Ref. \\
 & & & & & & & & or &    \\
& & l & b & (0.75--1) & (1--3) & ~~(3--10)~~ & (10--30) & Phase &    \\
&  & [deg] & [deg] & [MeV] & [MeV] & [MeV]  & [MeV] & &  \\
& & & & & & & & &   \\
\hline
\hline
 & & \multicolumn{2}{c|}{} &  \multicolumn{4}{c|}{} & & \\
3C273  & 0.158 & 290.0 &  64.4 & \multicolumn{2}{c|} {25.9$\pm$6.2} 
           & \multicolumn{2}{c|}{ $<$1.7} & VP 3  & (5) \\
(1226+023)& & & & \multicolumn{2}{c|}{(0.75--8 MeV)}
           & \multicolumn{2}{c|}{(8--30 MeV)}& &  \\
 & & & & 13.8$\pm$4.8 & 10.7$\pm$3.7 & 5.6$\pm$1.9 & $<$1.7 & VP 3 &  \\
 & & & & (0.75--1.25) & (1.25--3) & (3--8)  & (8--30) & &  \\
& & & & \multicolumn{2}{c|}{$<$19.2}
           & \multicolumn{2}{c|}{$<$2.5}& VP 11 & \\
 & & & & \multicolumn{2}{c|}{(0.75--8 MeV)}
           & \multicolumn{2}{c|}{(8--30 MeV)} & & \\
& & & & $<$8.5 & 14.5$\pm$2.2 & 3.9$\pm$0.9 & 0.6$\pm$0.3 & Phase I to III  
      & (6) \\
& & & &  & & & & &   \\
\hline
& & & &  & & & & &   \\
GROJ1224+2155 & & 255.1 & +81.7 & & & & & & \\
(PKS 1222+216) & & & &  $<$4.8 &  $<$5.4 & 2.6$\pm$0.9 
             & $<$0.9 & Phase I--III & (6) \\
& & & &  & & & & &   \\
\hline
& & & & & & & & &    \\
3C279 & & & &\multicolumn{2}{c|}{} & \multicolumn{2}{c|}{} & & \\
(1253--055) & 0.538 & 305.1 & 57.1 & \multicolumn{2}{c|}{$<$13.8}
            & \multicolumn{2}{c|}{2.9$\pm$0.9}
            & 3 & (5) \\
 & & & & \multicolumn{2}{c|}{(0.75--8 MeV)}
           & \multicolumn{2}{c|}{(8--30 MeV)} & & \\
 & & & & \multicolumn{2}{c|}{$<$12.1}& \multicolumn{2}{c|}{$<$3.2} 
 & 1  & \\
 & & & & \multicolumn{2}{c|}{(.75--8 MeV)}
           & \multicolumn{2}{c|}{(8--30 MeV)} & & \\
& & & & & & & & &   \\
& & & & $<$8.8 & 6.3$\pm$2.0 & $<$3.2 & 0.8$\pm$0.3 & Phase I to III & (6) \\ 
& & & & 5.9$\pm$3.0 & 6.9$\pm$2.4 & 1.5$\pm$1.0 & 1.0$\pm$0.3 & Phase I to IV & (9) \\ 
& & & & & & & & &   \\
\hline
Centaurus A & $\sim$3 Mpc & 309.5 & 19.4 & 16.4$\pm$8.1 & $<$17.6 & 4.0$\pm$2.4 & 1.6$\pm$0.8 & VP 12 & (7)  \\
NGC 5128  & & & & 31.0$\pm$ 12. & 16.4$\pm$ 11 & $<$9.2 & $<$5.8
 & VP 14$^{\rm a}$ & \\
& & & & 16.8$\pm$ 16.  & $<$28.6 & $<$15.8 & $<$ 2.9
  & VP 23$^{\rm b}$ & \\
 & & & & $<$20.1 & 30.3$\pm$13. & 7.1$\pm$6.2 & $<$3.9 & VP 27 & \\
  & & & & $<$20.5 &  $<$35.0 & $<$14.9 & $<$5.7 & VP 32 & \\
& & & & 14.1$\pm$5.2 &10.8$\pm$4.5 & 2.2$\pm$1.9 & $<$ 1.4
  & Phase I & \\
& & & &  & & & & & \\
& & & & $<$12.2 &  $<$14.3  & 8.8$\pm$2.6 & 1.3$\pm$0.9 & VP 207 & \\
& & & & 18.8$\pm$15. & 14.5$\pm$12. & 5.1$\pm$4.4 & $<$3.8 
  & VP  208 & \\
& & & & $<$23.3 & 21.3$\pm$8.3 & 10.8$\pm$3.1 & $<$1.7 & VP 215 +217 & \\
& & & & $<$11.9 & 5.8 $\pm$4.6 & 8.6$\pm$1.8 & $<$1.7 & Phase II & \\
& & & & & & & & &   \\
\hline\multicolumn{10}{r}{(Table 5 cont.)}\\
\end{tabular}

\clearpage

\begin{tabular}{||l|c|c|c|c|c|c|c|c|c||}
\hline\hline \noalign{\smallskip}
 & & \multicolumn{2}{c|}{} &  \multicolumn{4}{c|}{} & & \\
Source & Z-Value & \multicolumn{2}{c|}{Position} & \multicolumn{4}{c|}{Flux (10$^{-5}$ photon cm$^{-2}$ s$^{-1}$)}  & VP  & Ref. \\
 & & & & & & & & or &    \\
& & l & b & (0.75--1) & (1--3) & ~~(3--10)~~ & (10--30) & Phase &    \\
&  & [deg] & [deg] & [MeV] & [MeV] & [MeV]  & [MeV] & &  \\
& & & & & & & & &   \\
\hline
\hline
& & \multicolumn{2}{c|}{} &  \multicolumn{4}{c|}{} & & \\
Centaurus A ~~~~ & & & & 28.1$\pm$9.9 & $<$26.6 & $<$6.7  & 2.3$\pm$1.4~ 
  & VP 314  & \\ 
(NGC 5128)  & & & & $<$6.9 & 4.5$\pm$2.9 & $<$2.2 & $<$2.4 & VP 315 & \\
 & & & & $<$30.6 & $<$17.1 & $<$9.0 & $<$1.6 & VP 316 & \\
& & & & 24.1 $\pm$6.2 & $<$14.9 & $<$3.9 & $<$1.2
  & Phase III & \\
& & & &  & & & & & \\
& & & & 18.7$\pm$14. & $<$36.1 & $<$9.1  & $<$3.0  
  & VP 402.0 & \\
& & & & $<$28.7 & 26.6$\pm$12. & $<$8.9 & $<$4.2 &  VP 402.5 & \\ 
& & & & $<$13.5 & 18.6$\pm$7.7 & 3.6$\pm$2.7  & $<$1.3 &  VP 424.0 & \\
& & & & $<$15.2 & ~~21.9$\pm$5.8~~ & ~$<$4.5~ & $<$1.1 &  Phase IV & \\ 
& & & & & & & & & \\
& & & & 9.1$\pm$3.0 & 6.4$\pm$2.5 & ~2.1$\pm$1.0 & $<$0.6 & Phase I to IV  & \\    & & & & & & & & &   \\
\hline
& & & & & & & & & \\
PKS 1622--297 & 0.815 & 348.5 & 13.5 & $<$22.0
 & $<$13.5 & $<$6.1 & 3.3$\pm$0.7  & VP 421--423.5 & (8) \\
& & & & & & & & & \\
\noalign{\smallskip}\hline\hline
\end{tabular}  
\vskip6pt
{\bf Notes}
\vskip6pt
$^{\rm a}$Heavily influenced by Earth in field-of-view.

$^{\rm b}$Large data loss due to malfunctioning tape recorder.
\vskip6pt
{\bf References}
\vskip6pt
(1)  \cite{blom95a}.
 
(2) \cite{collmar97a}.
 
(3)  \cite{bloemen95}.

(4)  \cite{blom95b}.

(5)  \cite{williams95b}.

(6)  \cite{collmar96}.

(7)  \cite{steinle98}.

(8)  \cite{collmar97b}.

(9)  \cite{collmar97c}.

\clearpage


\landscape
\footnotesize

{\bf Table 6}: {Unidentified High-LatitudeSource}
\begin{flushleft}
\begin{tabular}{||l|c|c|l|l|l|l|l|l|c||}
\noalign{\smallskip}
\hline
\hline
& \multicolumn{2}{c|}{}  & \multicolumn{5}{c|}{} &  & \\
 & \multicolumn{2}{c|}{COMPTEL Position}
  & \multicolumn{5}{c|}{Flux (10$^{-5}$ cm$^{-2}$ s$^{-1}$)} & & \\
Source  & l & b & (0.75--1) &  (1--3) &   (3--10) &  
 (10--30)  &  Other Energy & VP/Phase
        &  Ref. \\
&[deg] &[deg] & [MeV]  & [MeV] &  [MeV] &  [MeV] & Ranges [MeV] &  &  \\
\hline
\hline
& & & & & & & & &  \\
GRO J 1753+57$^{\rm a}$ & 85.5 & 30.5 & $<$13.9 & $<$14.0 &
  $<$13.5 & $<$2.4 & & VP 2  & (1) \\
& & & $<$17.9 & $<$13.8 & $<$7.2 &
   $<$2.8 & & VP 9.5         & (6) \\
& & & $<$24.7 & 48.9$\pm$9.1 & $<$12.2 & $<$2.2 
   & & VP  201  & \\ 
& & & 18.6$\pm$9.9 & 28.8$\pm$9.4 & $<$12.1 &
  $<$4.9 & & VP 202  & \\ 
& & & $<$7.0 & $<$8.8 & 5.6$\pm$2.7 & $<$2.8 
  & & VP 203  & \\ 
& & & $<$9.4 & 11.0$\pm$5.1 & $<$8 &  $<$2.0 
  & & VP 212  & \\ 
\hline
& & & & & & & & &  \\
GRO J 1040+48 & 165 & 57 & & & & $<$0.56 & 19.8$\pm$3.2 (0.75--1.36) 
              & Phase I - II&  (2) \\
& & & & & & & 6.48$\pm$1.08(1.56--2.1 &  & \\
& & & & & & & 1.97$\pm$1.6(2.3--4.0 &  & \\
& & & & & & & $<$2.58 (4--10)&  & \\
& & & & & & & 24$\pm$3 (0.75--3 & Phase I - II & \\
& & & & & & & 7$\pm$4 (0.75--3) & Phase III - IV  & \\
\hline
& & & & & & & & &  \\
GRO J 1214+06 & 278.9 & +66.6 & $<$5.1 & $<$6.8 & 4.0$\pm$0.9 & $<$0.7 
              & & Phase I - III  & (3) \\
\hline
& & & & & & & & &  \\
Extended emission & 145$<$l$<$195 
& 35$<$b$<$65 & & & & &  150$\pm$10 (0.75--3) & Phase I to IV &
 (4) \\
from the HVC & & & & & & & & &\\
complexes  & & & & & & & & & \\
M and A area$^{\rm b}$ & & & & & & & & & \\
\hline
& & & & & & & & &  \\
Extended emission & 75$<$l$<$95 & 25$<$b$<$45 & & & & & 
    110$\pm$10(0.75-3) & Phase I to IV  & (5) \\
from the HVC & & & & & & & & & \\
complex C area$^{\rm c}$ & & & & & & & & & \\
\hline
\hline
\end{tabular}
\end{flushleft}
\vskip6pt
\noindent
$^{\rm a}$The emission cannot arise from a single source, but it can be modelled as a combination of emission from both GRO J1837+59 (a bright \\
\phantom{$^{\rm a }$}unidentified EGRET source), and the steep spectrum EGRET blazar QSO 1739+522. \\
\noindent
$^{\rm b}$Cloud region contains: 2EG J0917+4420, 2EG J0957+5515, GRO J1040+48.\\
\noindent
$^{\rm c}$Cloud region contains: GRO J1753+57, 2EG J1739+5152,
 2EG 1731+6007, 2EG J 1835+5913
\vskip6pt
{\bf References} 
\vskip6pt
(1)  \cite{williams95a}.

(2)  \cite{iyudin96}.

(3)  \cite{collmar96}.

(4)  \cite{blom97b}.

(5)  \cite{blom97a}.

(6)  \cite{williams99}.

\clearpage
\endlandscape

\landscape

\footnotesize
{\bf Table 7}: {Gamma-Ray Line Sources}
\vskip12pt
\begin{tabular}{||l|c|c|c|c|l|l|c||}             
\hline \hline \noalign{\smallskip} 
& \multicolumn{2}{c|}{} & & & & & \\
Name & \multicolumn{2}{c|}{Position} &  Line Energy & Flux 
    & VP or Phase & Note & Ref.  \\
 & l & b  &  & &   & & \\
   & [deg]  & [deg]  & [MeV] 
 & [10$^{-5}$ photons & & & \\
& &  & &  cm$^{-2}$ s$^{-1}$] & & & \\
  & & & & & & & \\
\hline 
\hline                                                    
& & & & & & & \\  
SN 1991T~~~~~~~~~~~~~~~ & ~~~~292.61~~~~ & ~~~~65.19~~~~ & 0.847 
         & $5.3 \pm 2.0 $ & VP 3.0, 11.0, 204.0, &  &   \\
\end{tabular}
\endlandscape

\landscape
\begin{tabular}{||l|c|c|c|c|l|l|c||}             
\hline \hline \noalign{\smallskip} 
& \multicolumn{2}{c|}{} & & & & & \\
Name & \multicolumn{2}{c|}{Position} &  Line Energy & Flux 
    & VP or Phase & Note & Ref.  \\
 & l & b  &  & &    & & \\
   & [deg]  & [deg]  & [MeV] 
 & [10$^{-5}$ photons & & & \\
& &  & &  cm$^{-2}$ s$^{-1}$] & & & \\
  & & & & & & & \\
\hline 
\hline
 & & & & & & &  \\ 
 Cygnus Region ~~& 73$<l<$93 & -7$<b<7$
              & 1.809 & 7.0 $\pm$ 1.4 & VP 2.0, 7.0, 34.0 & extended source & \\
 & & & & &  203.x, 212.0, 302.0, & & \\
 & & & & &  303.x, 318.1, 328.0, & & (5)\\
  & & & & & 331.x, 333.0         & & \\
 & & & & & & &  \\
\hline
 & & & & & & &  \\
Cygnus Superbubble & ~~73$<l<$87~~ & -7$<b<$+7
   & 1.809 & 4.6 $\pm$ 1.1 & see Cygnus region & extended source & (5) \\
 & & & & & & &  \\
\hline
& & & & & & &  \\
Carina Region & 286.5 & 0.5
   & 1.809 & 2.7 $\pm$ 0.6 & Phase I to IV/& point-like source & (4) \\
 & & & & & Cycle-5 & & \\
  & & & & & & &  \\ 
\hline
& & & & & & &  \\
Perseus/Cas-Tau~~~& 125$<$l$<$168 & -17$<$b$<$28 & 1.809 & 8.4$\pm$ 1 & 
Phase I to IV/ & low level extended & (4) \\
Region & & & & & Cycle-5 & emission &      \\ 
& & & & & & &  \\ 
\hline
& & & & & & &  \\
Neutron capture~~~ & ~~300~~~~~ &~~~~~ -30~~~~~ & 2.223 & 3.3$\pm$ 0.9 & 
VP 1.0 to  &  & (6) \\
source & & & & & VP 523.0  & white dwarf         & \\
& & & & & ~~~~~~~~         & RE J0317-853        &  \\
& & & & & ~~~~~~~~         & as possible source  &   \\ 
& & & & & ~~~~~~~~         & candidate           &  \\
\noalign{\smallskip}\hline\hline
\end{tabular}                                                      
\vskip12pt
{\bf References}
\vskip3pt
(1)  \cite{morris95}.

(2)  \cite{iyudin94}.

(3)  \cite{schoenf96}.

(4)  \cite{oberlack97}.

(5)  \cite{delrio96}.

(6)  \cite{mcconnell97a}.

(7)  \cite{dupraz97}.
 
(8) \cite{iyudin98}. 
\endlandscape

\clearpage


\landscape
\def\p{{$\pm$}}
\def\s{{$\sigma$}}

\footnotesize

{\bf Table 8}: {Gamma-Ray Burst Source Locations}
\begin{flushleft}
\begin{tabular}{||l|c|c|c|c|c|c|c|c|c|l||}
\noalign{\smallskip}
\hline
\hline
 & \multicolumn{3}{c|}{} & & & & \multicolumn{2}{c|}{} & &  \\
Burst  & \multicolumn{3}{c|}{COMPTEL Location} & Accum.
& (0.75--30) & Detect.
&\multicolumn{2}{c|}{Spectral Fit in Telescope Mode$^{\rm a}$}
       & Variability & References \\ 
   &  &  &  & Time  & [MeV]  
  & Signif. &  
    & & of Spect. in & \\ 
  & l & b &  Error & & Fluence  & & A & $\alpha$ 
  &  Single Det. & \\
  & & & Radius$^{b}$& & [10$^{-5}$ & &[ph cm$^{-2}$ s$^{-1}$ & & Mode & \\
  & [deg] & [deg] & [deg]  &  [s]  & erg cm$^{-2}$] & $\sigma$
  & MeV$^{-1}$] & & & \\
 & & & & &  & &  &  & & \\
\hline
\hline
 & & & & & & & & & & \\
\end{tabular}
\end{flushleft}
\endlandscape


\landscape




\footnotesize
{\bf Table 8}: {Gamma-Ray Burst Source Locations (cont.)}
\begin{flushleft}
\begin{tabular}{||l|c|c|c|c|c|c|c|c|c|l||}
\noalign{\smallskip}
\hline
\hline
 & \multicolumn{3}{c|}{} & & & & \multicolumn{2}{c|}{} & &  \\
Burst  & \multicolumn{3}{c|}{COMPTEL Location} & Accum.
& (0.75--30) & Detect.
&\multicolumn{2}{c|}{Spectral Fit in Telescope Mode$^{\rm a}$}
       & Variability & References~~~~~ \\ 
   &  &  &  & Time  & [MeV]  
  & Signif. &  
    & & of Spect. in & \\ 
  & l & b &  Error & & Fluence  & & A & $\alpha$ 
  &  Single Det. & \\
  & & & Radius & & [10$^{-5}$ & &[ph cm$^{-2}$ s$^{-1}$ & & Mode & \\
  & [deg] & [deg] & [deg]  &  [s]  & erg cm$^{-2}$] & $\sigma$
  & MeV$^{-1}]$ & & & \\
\hline
\hline
& & & & & & & & & & \\
GRB 940217 ~~~~ & 154.5 & -55.5 & 0.69 & 162.79 & 17.40& 18.8  & 0.38 $\pm$ 0.02
& 2.60 $\pm$ 0.07 & yes ($>$ 0.3 MeV) & 1, 2, 4, 10 \\
\hline
\hline
\end{tabular}
\end{flushleft}
\noindent
$^{a}$ Power-law fit: dn/dE=
A (ph cm$^{-2}$ s$^{-1}$ MeV$^{-1}$) $\cdot$ (E/1 MeV) $^{-\alpha}$\\
$^{b}$ The error radius is defined as the angular radius of a circle having the same area as the irregularly shaped COMPTEL 1$\sigma$ confidence region.\\
\vskip8pt
\noindent
{\bf References}
\vskip12pt
\begin{flushleft}
\begin{tabular}{llllll}
1. \cite{kippen95d}.  & 7. \cite{hanlon94a}.  & 13. \cite{bennett93}. &  19. \cite{connors93b}. & 25. \cite{schaefer94}.   & 31. \cite{kippen98b}.  \\ 
2. \cite{kippen95c}.  & 8. \cite{ryan94b}.    & 14. \cite{ryan93}.   &  20. \cite{connors93a}.    & 26. \cite{hanlon94b}. & 32. \cite{connors97}.   \\
3. \cite{greiner95}.  & 9. \cite{kippen94c}.  & 15. \cite{kippen93}. &  21. \cite{winkler92}.     & 27. \cite{share94}.   &                           \\
4. \cite{winkler95}.  & 10. \cite{kippen94a}. & 16. \cite{varendorff93}. &  22. \cite{varendorff92}.  & 28. \cite{hanlon95}.  &                            \\
5. \cite{kippen95b}.  & 11. \cite{kippen94b}. & 17. \cite{winkler93b}.   & 23. \cite{schoenf91}.     & 29. \cite{kippen96}.  &                            \\ 
6. \cite{kippen95a}.  & 12. \cite{ryan94a}.   & 18. \cite{winkler93a}.   & 24. \cite{collmar93}.     & 30. \cite{kippen98a}.  &                             \\
\end{tabular}
\end{flushleft}

\endlandscape

\clearpage



\landscape
\def\p{{$\pm$}}
\def\s{{$\sigma$}}

\footnotesize

{\bf Table 9}: {Solar Flare Detections}
\begin{flushleft}
\begin{tabular}{||l|c|c|c|c|c|c|c|c|c|c||}
\noalign{\smallskip}
\hline
\hline
& & & & & & & & & &  \\
COMPTEL    & TJD & GOES Start & GOES  & BATSE  &  BATSE  &  COMPTEL & COMPTEL & 
COMPTEL    & Peak BATSE & GRO mass \\ 
Flare No.  &     & time & Class  & Flare No.  & Trigger No.  & Int. time (s) & 
total cts &  Peak cts   & cts/area  &  g/cm$^{2}$  \\ 
& & & & & & & & & &  \\
\hline
\hline
 & & & & & & & & & &  \\
sf 910604-033651 & 8411  & 03:37 & X 12.0 & 387  & 262  &  900  & ${\ast}$
& ${\ast}$ &  N / A  &  39.5  \\
sf 910606-005900 & 8413  & 00:54 & X 12.0 & 388  & 281  & 6845  & 31198 
& 589       &  N / A  &  44.4      \\
sf910609-013433  & 8416  & 01:37 & X 10.0 & 419  & 299  & 5075  & 5880    
& 550       &  N / A  &  22.8  \\
\end{tabular}

\vskip6pt
{\bf Notes:}
\noindent
$\dagger$  ~~Marginal detection. \\
\hspace{1,1cm}$\ast$  ~~Detector saturation.
\end{flushleft}                                 
 \endlandscape

\clearpage


\footnotesize
{\bf Table 10}:
{Upper Limits to Galactic Black Hole Candidates} \\
\hspace{2cm}{$|b|<10^\circ$}                     \\
\begin{flushleft}
\begin{tabular}{||l|c|c|c|c|c|c|c|c||}
\hline\hline \noalign{\smallskip}
 & \multicolumn{2}{c|}{} &  \multicolumn{4}{c|}{} & & \\
Source &  \multicolumn{2}{c|}{Position} 
        & \multicolumn{4}{c|}{2$ \sigma$ Flux (10$^{-5}$ 
cm$^{-2}$ s$^{-1}$)} & Note  & Ref. \\
 & & & & & & & &    \\
 & l & b & (0.75--1) & (1--3) & (3--10) & (10--30)
    & &    \\
 & [deg] & [deg] & [MeV] & [MeV] & [MeV]  & [MeV] & &   \\
& & & & & & & &    \\
\hline
\hline                                         
& & & & & & & &  \\
GRO J1719-24 & 2.4 & 7.0 & $<$3.5 & 7.8$\pm$2.6 & $<$2.2 & $<$0.78 
 & a,b,c & (1) \\
\end{tabular}
\end{flushleft}

\clearpage

\begin{flushleft}
\begin{tabular}{||l|c|c|c|c|c|c|c|c||}
\hline \hline \noalign{\smallskip} 
 & \multicolumn{2}{c|}{} &  \multicolumn{4}{c|}{} & & \\
Source &  \multicolumn{2}{c|}{Position} 
        & \multicolumn{4}{c|}{2$ \sigma$ Flux (10$^{-5}$ 
cm$^{-2}$ s$^{-1}$)} & Note  & Ref. \\
 & & & & & & & &    \\
 & l & b & (0.75--1) & (1--3) & (3--10) & (10--30)
    & &    \\
 & [deg] & [deg] & [MeV] & [MeV] & [MeV]  & [MeV] & &   \\
& & & & & & & &    \\
\hline
\hline
& & & & & & & &    \\
GRS 1009--45 & 275.9 & 9.4  & $<$7.0 & $<$5.6~~ & $<$3.6~~ & $<$1.06    
 & a,c & (1) \\
\end{tabular}
\end{flushleft}

\clearpage

\begin{flushleft}
\begin{tabular}{||l|c|c|c|c|c|c|c|c||}
\hline \hline \noalign{\smallskip} 
 & \multicolumn{2}{c|}{} &  \multicolumn{4}{c|}{} & & \\
Source &  \multicolumn{2}{c|}{Position} 
        & \multicolumn{4}{c|}{2$ \sigma$ Flux (10$^{-5}$ 
cm$^{-2}$ s$^{-1}$)} & Note  & Ref. \\
 & & & & & & & &    \\
 & l & b & (0.75--1) & (1--3) & (3--10) & (10--30)
    & &    \\
 & [deg] & [deg] & [MeV] & [MeV] & [MeV]  & [MeV] & &   \\
& & & & & & & &    \\
\hline \hline
& & & & & & & &    \\
GRS  1734--292 & 358.84~~~~~ &  1.40 & $<$3.68~~ & $<$5.69~~ & $<$1.53 & $<$1.25  
 & a,c & (2) \\
& & & & & & & &    \\
\hline
& & & & & & & &    \\
 1E 1740.7--2942 & 359.1 &  -0.1  & $<$4.0 & $<$ 4.2 & $<$2.0 & $<$1.66  
 & a,c & (1) \\
 & & & $<$3.67 & $<$5.53 & $<$2.17 & 0.93$\pm$0.36
 & a,b,c & (2) \\
& & & & & & & &    \\
\hline
\hline   
\end{tabular}  
\vskip6pt
{\bf Notes}
\vskip6pt
\noindent
(a)  Model for diffuse galactic emission included in analysis. \\

(b)  A positive flux measurement does not 
     necessarily indicate  
     \phantom{(b)} a source detection (see ref. 1).  

(c) All upper limit fluxes are applied to data from Phase  
    \phantom{(c) }I~to~III.                                 

\vskip6pt
{\bf References}
\vskip6pt
(1) \cite{mcconnell96}.
 
(2) \cite{vandijk96}.

\end{flushleft}                                 
                                            

\clearpage


\landscape
\def\p{{$\pm$}}
\def\s{{$\sigma$}}

\footnotesize

{\bf Table 11a}: {COMPTEL 2$\sigma$ upper limits to AGN (EGRET sources and Seyferts)}
\begin{flushleft}
\begin{tabular}{||l|c|c|c|c|c|c|c|c|c|l||}
\noalign{\smallskip}
\hline
\hline
&\multicolumn{4}{c|}{}&\multicolumn{4}{c|}{}&\multicolumn{2}{c||}{}\\
Name  & \multicolumn{4}{c|}{Location} & \multicolumn {4}{c|}{Upper limit (10$^{-5}$ cm$^
{-2}$ sec$^{-1}$)}   &  Type& Other name\\ 
&RA &DEC &l&b& (0.75--1.0) & (1 - 3)  & (3 - 10)  & (10 - 30) &     &  \\ 
& [deg]  &  [deg]  &[deg] &[deg] & [MeV]       &  [MeV]   & [MeV]     & [MeV]  & & \\
\hline
\hline
 & & & & & & & & & & \\
0003 + 199  &  1.59  &  20.21 & 108.77  &  - 41.42  &  $<$ 4.24  &  $<$ 4.92 
& $<$ 5.03  &  $<$ 0.90  &  SY &  MRK 335   \\
0007 + 107  &  2.63  & 10.97  & 106.98  & - 50.63   & $<$ 3.70   & $<$ 5.12 
& $<$ 2.46  &  $<$ 0.89  & SY & IIIZW2        \\
0121 - 590  &  20.94 &  - 58.81  &  295.08  &  - 57.82  &  $<$ 4.67  
&  $<$ 7.72 &  $<$ 2.14  & $<$ 0.84 &  SY  &  FAIRALL 9  \\
\end{tabular}
\end{flushleft}
\endlandscape



\landscape
\def\p{{$\pm$}}
\def\s{{$\sigma$}}

\footnotesize

{\bf Table 11a}: {COMPTEL 2$\sigma$ upper limits to AGN (EGRET sources and Seyferts (cont.))}
\begin{flushleft}
\begin{tabular}{||l|c|c|c|c|c|c|c|c|c|l||}
\noalign{\smallskip}
\hline
\hline
&\multicolumn{4}{c|}{}&\multicolumn{4}{c|}{}&\multicolumn{2}{c||}{}\\
Name  & \multicolumn{4}{c|}{Location} & \multicolumn {4}{c|}{Upper limit (10$^{-5}$ cm$^{-2}$ sec$^{-1}$)}   &  Type& Other name  \\ 
&RA &DEC &l&b& (0.75--1.0) & (1 - 3)  & (3 - 10)  & (10 - 30) &     &  \\ 
& [deg]  &  [deg]  &[deg] &[deg] & [MeV]       &  [MeV]   & [MeV]     & [MeV]  & &       \\
\hline
\hline
 & & & & & & & & & & \\
0827 + 243  &  127.71 &  24.18  &   200.02   &  31.87    &  $<$ 1.92  
& $<$ 1.97  &  $<$ 1.05  & $<$ 1.10          &   2 EG    &               \\
0829 + 046  &  127.95 &  4.50   &   220.69   &  24.33    &  $<$ 2.13  
& $<$ 2.98  &  $<$ 3.52  & $<$ 1.56          &  2 EG     &              \\
0836 + 710  &  130.36 &  70.89  &   143.54   &  34.43    &  $<$ 2.89  
& $<$ 3.55  &  $<$ 2.39  & $<$ 0.66          &  2 EG     &              \\
\end{tabular}
\end{flushleft}
\endlandscape


\landscape
\def\p{{$\pm$}}
\def\s{{$\sigma$}}

\footnotesize

{\bf Table 11a}: {COMPTEL 2$\sigma$ upper limits to AGN (EGRET sources and Seyferts (cont.))}
\begin{flushleft}
\begin{tabular}{||l|c|c|c|c|c|c|c|c|c|l||}
\noalign{\smallskip}
\hline
\hline
&\multicolumn{4}{c|}{}&\multicolumn{4}{c|}{}&\multicolumn{2}{c||}{}\\
Name  & \multicolumn{4}{c|}{Location} & \multicolumn {4}{c|}{Upper limit (10$^{-5}$ cm$^{-2}$ sec$^{-1}$)}   &  Type& Other name   \\ 
&RA &DEC &l&b& (0.75--1.0) & (1 - 3)  & (3 - 10)  & (10 - 30) &     &  \\ 
& [deg]  &  [deg]  &[deg] &[deg] & [MeV]       &  [MeV]   & [MeV]     & [MeV]  & &\\
\hline
\hline
 & & & & & & & & & & \\
1460 - 076 &  212.24    & -7.88    &  333.88   &  50.28    &  $<$ 4.48  
& $<$ 7.51 &  $<$ 1.74  & $<$ 0.68 &  2 EG     &               \\
1410 - 029 &  213.31    & -3.21    &  339.15   &  53.81    &  $<$ 3.80  
& $<$ 10.68&  $<$ 1.69  & $<$ 0.69 &  SY       &  NGC 5506      \\
1415 - 253 &  214.49    & 25.14    &  31.96    &  70.50    &  $<$ 6.36  
& $<$ 14.14&  $<$ 4.49  & $<$ 1.20 &  SY       &  NGC 5548       \\
\end{tabular}
\end{flushleft}
\endlandscape



\landscape
\def\p{{$\pm$}}
\def\s{{$\sigma$}}

\footnotesize

{\bf Table 11a}: {COMPTEL 2$\sigma$ upper limits to AGN (EGRET sources and Seyferts (cont.))}
\begin{flushleft}
\begin{tabular}{||l|c|c|c|c|c|c|c|c|c|l||}
\noalign{\smallskip}
\hline
\hline
&\multicolumn{4}{c|}{}&\multicolumn{4}{c|}{}&\multicolumn{2}{c||}{}\\
Name  & \multicolumn{4}{c|}{Location} & \multicolumn {4}{c|}{Upper limit (10$^{-5}$ cm$^{-2}$ sec$^{-1}$)}   &  Type& Other name  \\ 
&RA &DEC &l&b& (0.75--1.0) & (1 - 3)  & (3 - 10)  & (10 - 30) &     &   \\ 
& [deg]  &  [deg]  &[deg] &[deg] & [MeV]       &  [MeV]   & [MeV]     & [MeV]  & &\\
\hline
\hline
 & & & & & & & & & & \\
2230 + 114 &  338.15    & 11.73    &  77.44   &  -38.58   &  $<$ 3.36  
& $<$ 8.06 &  $<$ 2.96  & $<$ 0.98 &  2 EG    &  CTA 102              \\
2233 - 263 &  338.94    & -26.05   &  27.14   &  -59.74   &  $<$ 3.89  
& $<$ 5.48 &  $<$ 5.74  & $<$ 1.34 &  SY      &  NGC 7314		\\
2251 + 158 &  343.49    & 16.15    &  86.11   &  -38.18   &  $<$ 4.37  
& $<$ 9.28 &  $<$ 6.63  & $<$ 1.53 &  2 EG    &  3C 454.3		\\
2300 + 086 &  345.82    & 8.87     &  83.10   &  -45.47   &  $<$ 3.52 
& $<$ 6.98 &  $<$ 4.25  & $<$ 1.17 &  SY      &  NGC 7469		\\
2302 - 090 &  346.23    & -8.73    &  64.08   &  -58.82   &  $<$ 3.46  
& $<$ 4.68 &  $<$ 5.78  & $<$ 1.30 &  SY      &  MCG-2-58-22 		\\
2315 - 426 &  349.60    & -42.37   &  348.08  &  -65.70   &  $<$ 5.81  
& $<$ 8.86 &  $<$ 5.27  & $<$ 1.30 &  SY      &  NGC 7582		\\
2356 + 196 &  359.69    & 19.92    &  106.38  &  -41.25   &  $<$ 3.47  
& $<$ 4.84 &  $<$ 4.12  & $<$ 0.91 &  2 EG    &				\\
& & & & &  & &  &  & & \\
\hline
\hline
\end{tabular}
\end{flushleft}
\endlandscape

\clearpage


\landscape
\def\p{{$\pm$}}
\def\s{{$\sigma$}}

\footnotesize

{\bf Table 11b}: {COMPTEL 2$\sigma$ upper limits to unidentified high-latitude  EGRET sources}
\begin{flushleft}
\begin{tabular}{||l|c|c|c|c|c|c|c|c|c||}
\noalign{\smallskip}
\hline
\hline
&\multicolumn{4}{c|}{}  &\multicolumn{4}{c|}{}  &      \\
Name  & \multicolumn{4}{c|}{Location} & \multicolumn {4}{c|}{Upper limit (10$^{-5}$ cm$^{-2}$ sec$^{-1}$)}   &  Type    \\ 
&RA &DEC &l&b& (0.75--1.0) & (1 - 3)  & (3 - 10)  & (10 - 30) &       \\ 
& [deg]  &  [deg]  &[deg] &[deg] & [MeV]  &  [MeV]   & [MeV]     & [MeV]  & \\
\hline
\hline
 & & & & & & & & &  \\
2EGJ0008 + 7307  & 2.09  & 73.13 & 119.77 & 10.52  & $<$ 5.29 & $<$ 4.48  
& $<$ 2.18 & $<$ 1.01 & 2 EG	\\
2EGJ0119 + 0312  & 19.97 & 3.21  & 136.77 & -58.90 & $<$ 4.63 & $<$ 5.36 
& $<$ 8.43 & $<$ 1.03 & 2 EG	\\
2EGJ0159 - 3557  & 29.85 & -35.95 & 248.55 & -73.08 & $<$ 9.06 & $<$ 11.08 
& $<$ 3.95 & $<$ 2.46 & 2 EG 	\\
\end{tabular}
\end{flushleft}
\endlandscape

\clearpage

\landscape
\def\p{{$\pm$}}
\def\s{{$\sigma$}}

\footnotesize

{\bf Table 11b}: {COMPTEL 2$\sigma$ upper limits to unidentified high-latitude  EGRET sources (cont.)}
\begin{flushleft}
\begin{tabular}{||l|c|c|c|c|c|c|c|c|c||}
\noalign{\smallskip}
\hline
\hline
&\multicolumn{4}{c|}{}  &\multicolumn{4}{c|}{}  &      \\
Name  & \multicolumn{4}{c|}{Location} & \multicolumn {4}{c|}{Upper limit (10$^{-5}$ cm$^{-2}$ sec$^{-1}$)}   &  Type    \\ 
&RA &DEC &l&b& (0.75--1.0) & (1 - 3)  & (3 - 10)  & (10 - 30) &       \\ 
& [deg]  &  [deg]  &[deg] &[deg] & [MeV]  &  [MeV]   & [MeV]     & [MeV]  & \\
\hline
\hline
 & & & & & & & & &  \\
2EGJ1220 - 1510  & 185.00  & -15.17& 291.77 & 47.02 & $<$ 3.76 & $<$ 2.90  
& $<$ 5.12 & $<$ 1.53 & 2 EGS	\\
2EGJ1233 - 1407  & 188.27  & -14.13& 296.20 & 48.50  & $<$ 4.94 & $<$ 2.80 
& $<$ 4.19 & $<$ 0.72  & 2 EG	\\
2EGSJ1236 - 0416 & 189.00  & -4.27 & 295.58 & 58.38  & $<$ 2.28 & $<$ 3.73 
& $<$ 2.41 & $<$ 0.90 & 2 EGS	\\
\end{tabular}
\end{flushleft}
\endlandscape

\clearpage

\landscape
\def\p{{$\pm$}}
\def\s{{$\sigma$}}

\footnotesize

{\bf Table 11b}: {COMPTEL 2$\sigma$ upper limits to unidentified high-latitude  EGRET sources (cont.)}
\begin{flushleft}
\begin{tabular}{||l|c|c|c|c|c|c|c|c|c||}
\noalign{\smallskip}
\hline
\hline
&\multicolumn{4}{c|}{}  &\multicolumn{4}{c|}{}  &      \\
Name  & \multicolumn{4}{c|}{Location} & \multicolumn {4}{c|}{Upper limit (10$^{-5}$ cm$^{-2}$ sec$^{-1}$)}   &  Type    \\ 
&RA &DEC &l&b& (0.75--1.0) & (1 - 3)  & (3 - 10)  & (10 - 30) &       \\ 
& [deg]  &  [deg]  &[deg] &[deg] & [MeV]  &  [MeV]   & [MeV]     & [MeV]  & \\
\hline
\hline
 & & & & & & & & &  \\
2EGJ2006 - 2253  & 301.62  & -22.90 & 19.32 & -26.17 & $<$ 2.95 & $<$ 3.68  
& $<$ 3.88 & $<$ 0.77 & 2 EG	\\
2EGJ2027 + 1054  & 306.79  & 10.91  & 54.24 & -15.55 & $<$ 2.39 & $<$ 8.02 
& $<$ 4.73 & $<$ 1.01 & 2 EG	\\
2EGJ2243 + 1545  & 340.82  & 15.75  & 83.17 & -37.04 & $<$ 3.89 & $<$ 7.42 
& $<$ 4.96 & $<$ 1.64 & 2 EG	\\
2EGSJ2322 - 0321 & 350.60  & -3.36  & 77.24 & -58.10 & $<$ 3.68 & $<$ 4.76 
& $<$ 3.27 & $<$ 1.27 & 2 EGS	\\
2EGJ2354 + 3811 & 358.56 & 38.19    & 110.73& -23.32 & $<$ 4.41 & $<$ 5.41 
& $<$ 3.42 & $<$ 0.93 & 2 EG	\\
& & & & &  & &  &  &  \\
\hline
\hline
\end{tabular}
\end{flushleft}
\endlandscape

\clearpage


\landscape
\footnotesize
{\bf Table 12}: {Upper Limits to Possible Sources of
Gamma-Ray Line Emission}
\vskip12pt
\begin{tabular}{||l|c|c|c|c|l|c|c||}
\hline \hline \noalign{\smallskip}
& \multicolumn{2}{c|}{} & & & & & \\
Name & \multicolumn{2}{c|}{Position} &  Line Energy & Flux (2 $\sigma$)
    & ~~~~~Observation & Note & Ref.  \\
 & l & b  &  & & ~~~~~Period   & & \\
   & [deg]  & [deg]  & [MeV] 
 & [10$^{-5}$ photons & & & \\
& &  & &  cm$^{-2}$ s$^{-1}$] & & & \\
& & & & & & & \\
\hline 
\hline
& & & & & & & \\
SN 1993J & 142.15$^\circ$ & 40.92$^\circ$ & 0.847 
         & $<$6.2 & VP 227.0, 228.0, 319.0, &  &   \\
  &  & & 1.238 & $<$2.9 &  319,0, 319.5     & (a) &  (2) \\
 &  &  & &  & & &  \\
\hline
  &  &  & &  & & &  \\                  
Kepler SN (1604) & 4.52 & +6.84 & 1.157  &  $<$1.43 & Phase I to III & & (1) \\
 &  &  & &  & & &  \\ 
\hline
  &  & &  & & & &  \\ 
Tycho SN (1572) & 120.09 & +1.42 & 1.157  &  $<$1.78 & Phase I to III & & (1) \\
 &  &  & &  & & &  \\ 
\hline
  &  & &  & & & &  \\ 
3C 58    & 130.72   & +3.08 & 1.157  &  $<$1.78 & Phase I to III &  & (1) \\
 & & & &  & & &  \\ 
\hline
  & & &  & & & &  \\ 
Lupus SN (1006) & 327.51 & +14.69 & 1.157  &  $<$1.50 & Phase I to III & &(1) \\
 & & & &  & & &  \\ 
\hline
 & & & & & & &  \\ 
Novae (general) & & & 1.275 MeV  & $<$3 
  & April 1991--Aug 1993 & & (4)  \\
 & & & & & & &  \\
\hline
& & & & & &  & \\
Nova Cen 1991 & 309$^\circ$ & -1.04$^\circ$ & 1.275 
& $<$4.0 & VP 12.0, 14.0, 23.0, 27.0,   &      &    \\ 
& & & & & 208.0, 215,0, 217.0,           & (b)  & (4) \\
& & & & &  314.0, 315.0, 316.0          &      &      \\
& & & & & &  & \\
\hline
& & & & & &  & \\
Nova Her 1991 ~~~~~ & 43.3$^\circ$ & 6.6 $^\circ$ & 1.275 MeV 
& $<$3.3  & VP 20.0, 328.0, 331.x,        & & \\
	  & & & & &  333.0 & (b) & (4) \\
& & & & & &  & \\
\hline\multicolumn{8}{r}{(Table 12 cont.)}\\
\end{tabular}
\endlandscape

\clearpage

\landscape
\begin{flushleft}
\begin{tabular}{||l|c|c|c|c|l|c|c||}
\hline \hline \noalign{\smallskip} 
& \multicolumn{2}{c|}{} & & & & & \\
Name & \multicolumn{2}{c|}{Position} &  Line Energy & Flux (2 $\sigma$)
    & ~~~~~Observation & Note & Ref.  \\
 & l & b  &  & & ~~~~~Period   & & \\
   & [deg]  & [deg]  & [MeV] 
 & [10$^{-5}$ photons & & & \\
& &  & &  cm$^{-2}$ s$^{-1}$] & & & \\
& & & & & & & \\
\hline 
\hline
& & & & & &  & \\
Nova SGR 1991~~~~~& ~ 0.18$^{\circ}$ ~ & -6.94$^\circ$ & ~ 1.275 ~ &  $<$6.2 &  5.0, 210.0, 214.0, &       &       \\
& & & & &  219.4, 223.0, 226.0,   &  &  \\
& & & & &  231.0, 229.x, 232.x,   &  &  \\
& & & & &  302.3, 223.0, 324.0,   & (b) & (4) \\
& & & & &  330.0, 332.0, 334.0,   &  &  \\
& & & & &  336.5, 338.0           &  &  \\
&  & &  &  & & &   \\
\hline
&  & &  &  & & &   \\
Nova Sct 1991 &  \ \ 25.1$^\circ$ & \ -2.80$^\circ$ & 1.275  
              & $<$3.6 & 7.5, 13.0, 20.0,    &  & \\
             & & & & &  231.0, 229.x, 324.0,  & (b) & (4) \\
	     & & & & &  330.0, 332.0, 334.0  &  &   \\
&  & &  &  & & &   \\
\hline
&  & &  &  & & &   \\
Nova Pup 1991 ~~~~ & ~ 252.7$^\circ$ ~ & ~ -0.72$^\circ$ ~ & 1.275 
              & $<$ 5.5 & 41.0, 44.0, 230.0,  & &  \\
                & & & & & 301.0, 338.5       & (b) & (4) \\
&  & &  &  & & &   \\
\hline
&  & &  &  & & &   \\
Nova Cyg 1992 & 89.14$^\circ$ & 7.82$^\circ$ & 1.275 
              & $<$2.3 & 34.0, 203.x, 212.0, & (b) & (4)   \\
             & & & &   & 302.x, 318.1 & &  \\
& & & &  & & &  \\
\hline
& & & &  & & &  \\
Nova Sco 1992 & 343.8$^\circ$ & -1.61$^\circ$ & 1.275  
              & $<$5.9 & 27.0, 210.0, 214.0, & & \\
               & & & & & 219.4, 223.0, 226.0 & & \\
	       & & & & & 229.x, 232.x, 302.3 & (b) & (4) \\
	       & & & & & 323.0, 336.5, 338.0 & & \\
&  & &  &  & & &  \\
\hline\multicolumn{8}{r}{(Table 12 cont.)}\\
\end{tabular}
\end{flushleft}
\endlandscape

\clearpage

\landscape
\begin{flushleft}
\begin{tabular}{||l|c|c|c|c|l|c|c||}
\hline \hline \noalign{\smallskip} 
& \multicolumn{2}{c|}{} & & & & & \\
Name & \multicolumn{2}{c|}{Position} &  Line Energy & Flux (2 $\sigma$)
    & ~~~~~Observation & Note & Ref.  \\
 & l & b  &  & & ~~~~~Period   & & \\
   & [deg]  & [deg]  & [MeV] 
 & [10$^{-5}$ photons & & & \\
& &  & &  cm$^{-2}$ s$^{-1}$] & & & \\
  & & & & & & & \\
\hline 
\hline
& & & &  & & &  \\
Nova Sgr 1992--1 ~~~~~ & ~ 4.75$^\circ$ ~ & ~ -2.0 $^\circ$ ~ & 1.275  
              & $<$6.0 &  210.0, 214.0, 219.4,   & & \\
\end{tabular}
\end{flushleft}
\endlandscape

\clearpage

\landscape
\begin{flushleft}
\begin{tabular}{||l|c|c|c|c|l|c|c||}
\hline \hline \noalign{\smallskip} 
& \multicolumn{2}{c|}{} & & & & & \\
Name & \multicolumn{2}{c|}{Position} &  Line Energy & Flux (2 $\sigma$)
    & ~~~~~Observation & Note & Ref.  \\
 & l & b  &  & & ~~~~~Period   & & \\
   & [deg]  & [deg]  & [MeV] 
 & [10$^{-5}$ photons & & & \\
& &  & &  cm$^{-2}$ s$^{-1}$] & & & \\
  & & & & & & & \\
\hline 
\hline
&  & &  &  & & &  \\
HB 21 ~~~~~~~~~~~~ & ~ 88.86 ~  & ~ 4.80 ~ & 1.809  &  $<$2.30  & Phase I to III ~~~~~~  & (d) & (6)   \\
&  & &  &  & & &  \\
\hline
&  & &  &  & & &  \\
Mon Nebula ~~~~~~~ & ~ 204.96 ~ & ~ 0.45 ~ & 1.809  &  $<$1.60  & Phase I to III ~~~~~~
   & (c) & (6) \\
&  & & &  & & &  \\
\hline
&  & & & & & &  \\
Lupus Loop & 329.67 &  15.97 & 1.809  & $<$1.40  & Phase I to III
& (e) & (6) \\
& & & & & & & \\
\hline
& & & & & & & \\
Orion & 209.0$^\circ$ & -19.4$^\circ$ & 1.809
& $<$1.7 & VP 1.0 & &  \\
& & & & &  VP 2.5 (solar mode) & &   \\
& & & & &  VP  221.0 &  & (3) \\
& & & & &  VP 337.0 &  &    \\
& & & & & & & \\
\hline
& & & & & & & \\
Cyg X-1 & 71.3    &  +3.1  & 2.223  & $<$1.53  & VP 1--523 & & (5) \\
&  & & & & & &  \\
\hline
& & & & & & & \\
Sco X-1 & 359.09  & 23.78  & 2.223  & $<$ 1.7  & Phase I to IV / & & (5) \\
        &         &        &        &          & Cycle-5         & &     \\
& & & & & & & \\
\hline
& & & & & & & \\
Cyg X-3 &  79.85  & 0.70   & 2.223  & $<$ 1.2  & Phase I to IV / & & (5)  \\
        &         &        &        &          & Cycle-5         & &      \\
& & & & & & & \\
\hline
& & & & & & & \\
4U 1916-05 & 31.38 & -8.22  & 2.223  & $<$ 1.2  & Phase I to IV / & & (5)  \\
           &      &        &        &          & Cycle-5         & &      \\
& & & & & & & \\
\hline\multicolumn{8}{r}{(Table 12 cont.)}\\
\end{tabular}
\end{flushleft}
\endlandscape

\clearpage

\landscape
\begin{flushleft}
\begin{tabular}{||l|c|c|c|c|l|c|c||}
\hline \hline \noalign{\smallskip} 
& \multicolumn{2}{c|}{} & & & & & \\
Name & \multicolumn{2}{c|}{Position} &  Line Energy & Flux (2 $\sigma$)
    & ~~~~~Observation & Note & Ref.  \\
 & l & b  &  & & ~~~~~Period   & & \\
   & [deg]  & [deg]  & [MeV] 
 & [10$^{-5}$ photons & & & \\
& &  & &  cm$^{-2}$ s$^{-1}$] & & & \\
  & & & & & & & \\
\hline 
\hline
& & & & & & & \\
4U 1626-67 ~~~~~~~~~~~& 321.79~& -13.09~& 2.223  & $<$ 1.7  & Phase I to IV / ~~~~~& & (5)\\
           &       &       &        &          & Cycle-5         & &       \\
& & & & & & & \\
\hline
& & & & & & & \\
4U 1820-30 ~~~~~~~~~~ &   2.79& -7.91 & 2.223  & $<$ 1.1  & Phase I to IV / ~~~~~& & (5)\\
           &       &       &        &          & Cycle-5         & &         \\
& & & & & & & \\
\hline\hline 
\end{tabular}                                                      
\vskip6pt
{\bf References}
\vskip6pt
(1) \cite{dupraz97}.
 
(2) \cite{lichti96}.
 
(3) \cite{oberlack95}.

(4) \cite{iyudin95}.

(5) \cite{mcconnell97b}.

(6) \cite{knoed96}.

\vskip6pt 
\vskip12pt
 (a) Only VP 227/228 used for upper flux limit.

 (b) Only observations up to VP 232 used for upper flux limit.

 (c) radius 2$^\circ$

 (d) radius 1$^\circ$

 (e) radius 1.5$^\circ$
\end{flushleft}                                  
\endlandscape                                            

\clearpage


\landscape
\footnotesize

{\bf Table 13}: {Summary of Most Significant COMPTEL Source Detections.}
\begin{flushleft}
\begin{tabular}{||l|c|l||}
\noalign{\smallskip}
\hline
\hline
  &  &                                       \\
 \ {\bf Type of Source}  & {\bf Number of} & \ {\bf Comments}   \\
                                         &  {\bf Sources}   &               \\
  &  &                                          \\
\hline
\hline
  &  &                                           \\
{\bf Spin-Down Pulsars:}  &    3     &  Crab, Vela, PSR B1509-58.  \\
  &  & \\
\hline
  &  &  \\
{\bf Stellar Black Hole} &    2     &  Cyg X-1, Nova Persei 1992 (GRO J0422+32).  \\
{\bf Candidates:}        &          &                         \\
  &  &   \\
\hline
  &  &   \\
{\bf Supernova Remnants:} &   1  &  Crab nebula.            \\
(Continuum Emission)  &    &                       \\
  &  &    \\
\hline
\  &  &    \\
 {\bf Active Galactic Nuclei:}   &     10     & CTA 102, 3C 454.3, PKS 0528+134, GRO J 0516-609, PKS 0208-512, 3C 273,                                 \\
                           &            &   PKS 1222+216, 3C 279, Cen A, PKS 1622-297. \\
\  &  &    \\ 
\hline
  &  &     \\
{\bf Unidentified Sources:}      &          &                           \\
$\bullet$ $\mid$b$\mid$ $<$ 10$^{\circ}$    &     4   &  GRO J1823-12, GRO J2228+61 (2CG 106+1.5),  GRO J0241+6119 (2CG 135+01), \\
                            &           &  Carina/Vela region (extended).  \\
$\bullet$ $\mid$b$\mid$ $>$ 10$^{\circ}$    &     5   &  GRO J1753+57 (extended), GRO J1040+48, GRO J1214+06, \\
                          &           &                HVC complexes M and A area (extended), HVC complex C (extended). \\ 
   &  &       \\
\hline
   &  &         \\
{\bf Gamma-Ray Line Sources:}     &           &            \\
$\bullet$ 1.809 MeV ($^{26}$Al)&  3     &  Cygnus region (extended), Vela region  (extended, may include RX J0852-4621),                                                            \\ 
                           &            & Carina region.  \\
$\bullet$ 1.157 MeV ($^{44}$Ti)&    2  &  Cas A, RX J0852-4621 (GRO J0852-4642).   \\
$\bullet$  0847 and 1.238 MeV ($^{56}$Co)    &  1  &  SN 1991T.        \\
$\bullet$  2.223 MeV (n-capture)         &  1  &  GRO J0317-853.        \\
  &  &     \\
\hline
  &  &     \\
{\bf Gamma-Ray Burst Sources:}    &     31 &  Location error radii vary from 0.34$^{\circ}$ to 2.79$^{\circ}$ (mean error radius: view 1.13$^{\circ}$).                  \\   
(within COMPTEL field-of-   &        &                                     \\
up to Phase IV/Cycle-5)     &        &                                      \\
 &   &      \\
\hline
\hline
\end{tabular}
\end{flushleft}

\endlandscape

\end{document}